\documentclass{JINST}
\usepackage{amsmath,amsfonts,amssymb,fontenc,times,mathptmx,graphicx,amsthm}
\title{Optimizing momentum resolution with a new fitting method for silicon-strip detectors}
\author{Gregorio Landi$^a$\thanks{Corresponding
author.}~,   Giovanni E. Landi$^b$\\
\\
\llap{$^a$} Dipartimento di Fisica e Astronomia,
Universita' di Firenze\\
Largo E. Fermi 2 (Arcetri) 50125 Firenze Italy\\
and INFN, Sezione di Firenze,
Firenze,Italy\\
E-mail: \email{Gregorio.Landi@fi.infn.it}\\
\\
\llap{$^b$} ArchonVR S.a.g.l.,\\
Via Cisieri 3,\\
6900 Lugano, Switzerland.\\}

\abstract{
A new fitting method is explored for momentum reconstruction. The tracker model
reproduces a set of silicon micro-strip detectors in a constant magnetic field.
The new fitting method gives substantial increases of momentum
resolution respect to standard fit. The key point is the use of a realistic
probability distribution for each  hit (heteroscedasticity).
Two different methods are used for the fits, the first method introduces
an effective variance as weight for each hit, the
second method uses the search of the maximum likelihood.
The tracker model is similar to the PAMELA tracker with
its two sided detectors. The two detector sides have very different
properties and quality. Each side is simulated as momentum
reconstruction device.  One of the two is similar to silicon micro-strip
detectors of large use in running experiments.
Two different position reconstructions are used for the standard fits, the
$\eta$-algorithm (the best one) and the two-strip center of gravity.
The gain obtained in momentum resolution is measured
as the virtual magnetic field
and the virtual signal-to-noise ratio required by the two standard fits to reach an overlap
with the best of two new methods. For the low noise side, the virtual magnetic
field must be increased 1.5 times respect to the real field to reach
the overlap and 1.8 for the other. For the high noise side, the
increases must be 1.8 and 2.0.  The virtual signal-to-noise ratio has to be
increased by 1.6 for the low noise side and 2.2 for the high noise side ($\eta$-algorithms).
Changes of the signal-to-noise ratio has no effect on the fits with the center
of gravity as position-algorithm. The momentum resolution is simulated even in function of the
number N of the detection layers. A very rapid linear increase with N is observed for
our two methods, the two standard fits has the usual grow as $\sqrt \mathrm{N}$.
Other interesting effects are  obtained selecting tracks with good or excellent hits.}


\begin{document}

\section{Introduction }

The momenta of the charged particles are fundamental pieces of
information for the event reconstructions in high energy
particle physics. Very complex instruments (trackers) have
been developed for this task where the key element is
a uniform (or near to) magnetic field.
In a homogeneous magnetic field a free charged particle
describes a helicoidal path, the radius of the helix is proportional to
the momentum component orthogonal to the magnetic field.
The chirality of the helix is given by the sign of the
charge. The precise reconstructions of these non linear paths
allow the trackers to measure the particle momenta and to fix
the charge signs.
Various effects  deviate the particle tracks from their theoretical
forms and refined methods must be introduced to account for these
deviations and limit the measurement degradations.
Among the degrading effects we quote the multiple (Coulomb) scattering, the energy loss,
the inhomogeneities of the magnetic field and the
systematic and statistical errors of the positioning algorithms.
Recent papers~\cite{Berger, Blobel} (to mention the most recent ones
of a very large set) are addressed to the first three effects.
The handling of the full non linearity of the particle path is
clearly described in ref.~\cite{Karimaki}.

This work is addressed to an accurate study of the statistical effects
in the positioning algorithms of minimum ionizing particle (MIP)
and how to reduce at minimum their influence in the
momentum reconstruction. For this task, realistic probability density
functions (PDFs; probability distributions in the physicist 
jargon) for the errors of the positioning algorithms must
be used.
This strategy allows to go beyond
the results of the standard least squares method or other fitting method that neglects
the hit-statistical properties, or the heteroscedasticity (as it
called in literature).
For our task, we will simulate momentum values
where the multiple scattering is unimportant, similarly for the energy
loss (principally for electrons) and the magnetic inhomogeneities.
The systematic errors in the positioning algorithms are unavoidable and
must be accounted properly. The center of gravity
(COG in the following) as a positioning algorithm
were detailed discussed in ref.~\cite{landi01,landi02} where exact analytical forms
were demonstrated for its systematic error. Due to its (apparent) simplicity the COG
is one of the most used positioning algorithms: $x_g=\sum_j x_j E_j/\sum_j E_j $, where
$E_j$ and $x_j$ are the signals and the positions used. Even if a
single name is used for the COG, slight different version are in use. The principal differences are in
number of signal used, each selection implies different analytical
properties and different systematic errors. The differences of each COG-version
are quite evident in our approach, the realistic PDFs show very different analytical
forms and statistical properties. To keep track of these differences we will indicate
with COG$_n$ a COG algorithm with n-strips. Due to these different forms and properties,
a rigid distinction must be exercised
among the various COG$_n$ algorithms and use the best one for each condition.
The developments of ref.~\cite{landi01} and the mathematical approaches
explored there for the COG study are very complex, but
these complexities are typical of the fight against the
systematic errors. For example, Gauss demonstrated what he called
{\em"Theorema Egregium"} for the systematic errors of his topographic maps,
and the recommendations on the systematic error suppression
are always preliminary statements in his papers on least squares~\cite{gauss}.
With heteroscedasticity as an essential
assumption, ref~\cite{gauss} deviates largely from
the standard expositions of the least squares,
those simplifications (often presented as Gauss-Markov theorem)
surely generate many suboptimal fits.

An interesting achievement, to suppress the COG$_2$ systematic error,
was reported in ref.~\cite{belau} i.e. the
$\eta$-algorithm. A preliminary assumption of the method is
the uniform "illumination" of the strips by MIP signals from parallel
incidence. In this case the non uniform
distribution of the COG$_2$, given by systematic error, can be reduced
to a uniform one. As clearly stated by the
authors, the derivation of ref.~\cite{belau} is limited to symmetric signal
distributions collected by the two (adjacent) leading strips of a cluster. It is
evident that for inclined tracks or in presence of a magnetic field the symmetry,
if present at orthogonal incidence, is surely destroyed. To overcame
these limitations, the analytic methods developed in ref.~\cite{landi01}
was essential. It
was used in ref.~\cite{landi03} to  generalize the $\eta$-algorithm beyond the two
strip limitation and to extract the required corrections for non symmetric configurations.
The set of generalizations will be indicated as $\eta_n$-algorithm, where the index
$n$ is connected to the number of strips contained in the COG$_n$ algorithm
from which the $\eta$ algorithm is derived.
Here, we will use always
the COG$_2$ and the $\eta_2$-algorithm.
At our orthogonal incidence, the signals collected by the two leading strips
suffice for the reconstructions.

The fit of straight tracks was the first problem we faced
with this new fitting method, even if the straight tracks
are not the principal aim in high energy physics.  The results
obtained were described in ref.~\cite{landi05}.
The limitation to straight tracks was mainly due to the
complexity of the method we applied for the first time. Hence, working with
two parameters, the debugging and testing of the application
can be followed on a surface. Furthermore, the data, elaborated in our approach, were collected
in a CERN test  beam~\cite{vannu} in the absence of a magnetic field.
The detectors, we used, were few samples of double-sided silicon microstrip
detector~\cite{aleph,L3} as those composing the PAMELA tracker~\cite{PAMELA}.
The results of ref.~\cite{landi05} showed a drastic improvement of
the fitted track parameters respect to the results of the least squares methods.
We observed excellent reconstructions even in presence of very noisy hits, often
called {\em outliers}, that generally produce bad fits. This achievement
is almost natural for  the heavy tails of our non gaussian PDFs . On the
contrary, the least squares method, being strictly optimum for a gaussian PDF,
must be modified in a somewhat arbitrary way to handle these pathological (for a
gaussian PDF) hits, often with scarce or no result.

We have to recall that the perception of a rough handling of the
hit PDFs is well present in literature, and Kalman-filter
modifications are often studied to accept extended deviations
from a pure Gaussian model. These extensions use linear
combinations of different Gaussians~\cite{fru}, the number of
them is limited to avoid the intractability of the equations.
The unknown parameters are selected from an optimization of
the fit results. In late sense, our schematic approximations could be
reconnected to those extensions. In fact, to speed the
convergence of the maximum likelihood search, we calculate an effective
variance for each hit and use this variance in a weighted
least squares.

The confidence gained in ref.~\cite{landi05} with straight tracks
allow us to deal with more complex and appealing tasks,
i.e. the reconstruction of tracks in a magnetic field and
the measure of their momenta.
To face this extension with the minimum modifications of our previous work,
and saving a backward compatibility,
we will  utilize the simulated data and parameters
used in ref.~\cite{landi05} thus this reference will be often quoted as
a repository of details not reported here.
Now those data are adapted to the PAMELA geometry and to a constant magnetic
field of $0.44\, T$ near to the average field of the PAMELA tracker.
This relatively low magnetic field introduces small
modifications of the parameters obtained in its absence. The average
signal distribution of a MIP is slightly
distorted by a Lorentz angle, that, in the present conditions, is about
$0.7^\circ$. Around this incidence angle, the average signal
distribution of a MIP is practically identical to that at
orthogonal incidence in the absence of a magnetic field. Thus,
without further notice, we will assume a rotation of the detectors of
their Lorentz angle respect to our MIP direction.
These assumptions allow to reutilize our previously generated data
and to simulate high momentum tracks for
each side of the double sided detectors. In the PAMELA tracker, the
low noise side of the detector (junction side) is used for
momentum measurements. This side has an excellent resolution due to its special
structure. A strip each two is connected to the read out system,
the unconnected strip distributes the incident signal on the nearby
strips optimizing the detector resolution (the floating strip side).
The other side (the
ohmic side) is devoted to the reconstruction of the straight part of a track,
it has the strips oriented perpendicularly to the junction
side. Each strip is connected to the readout system and has, by
the complexity of construction, an higher noise and a small, if any,
signal spread to nearby strips. For this last characteristic,
the ohmic side responds in a way similar to the types
of microstrip arrays used in the large trackers of the CERN
LHC~\cite{ALICE,ATLAS,CMSa}.
The simulations on this side, as bending side, can give
a glimpse of our approach for other type of trackers even for
the (small angle stereo) double-sided detectors of the
ALICE~\cite{ALICE} experiment. For these reasons, this side
will be indicated as normal noisy side.

In section 2, few details of this new fitting method will be recalled with
a derivation of the general probability distribution.
Sections 3 and 4 are devoted to the momentum PDF for
the each side of this two sided detectors.
In section 5 the results of the simulations will be
discussed and related to physical properties of the detection
system. Section 6 summarizes the results. A partial preliminary
version of this paper was reported in arXiv 1606.03051.

\section{Details of the method}

\begin{figure}[ht]
\begin{center}
\includegraphics[scale=0.5]{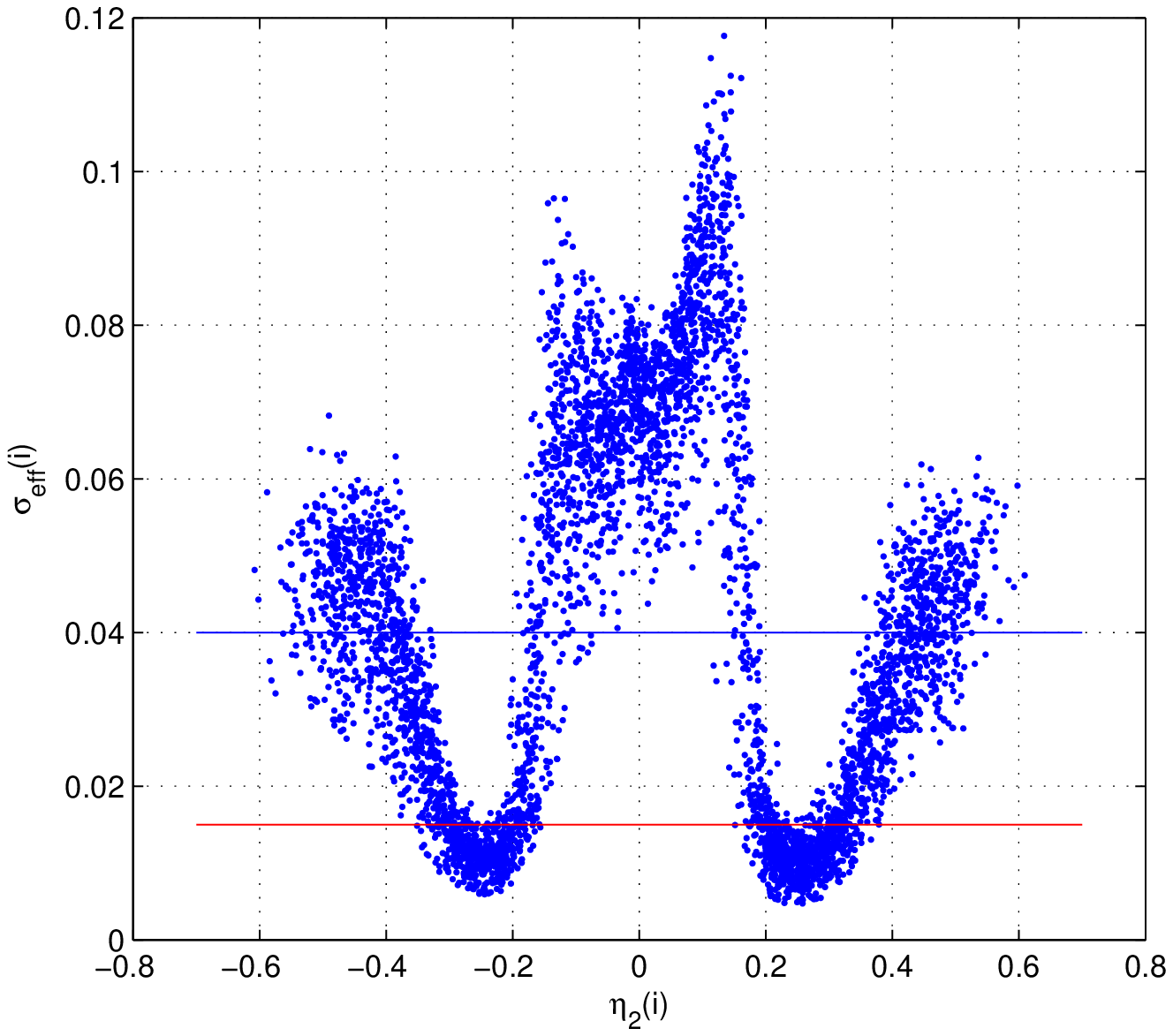}
\includegraphics[scale=0.5]{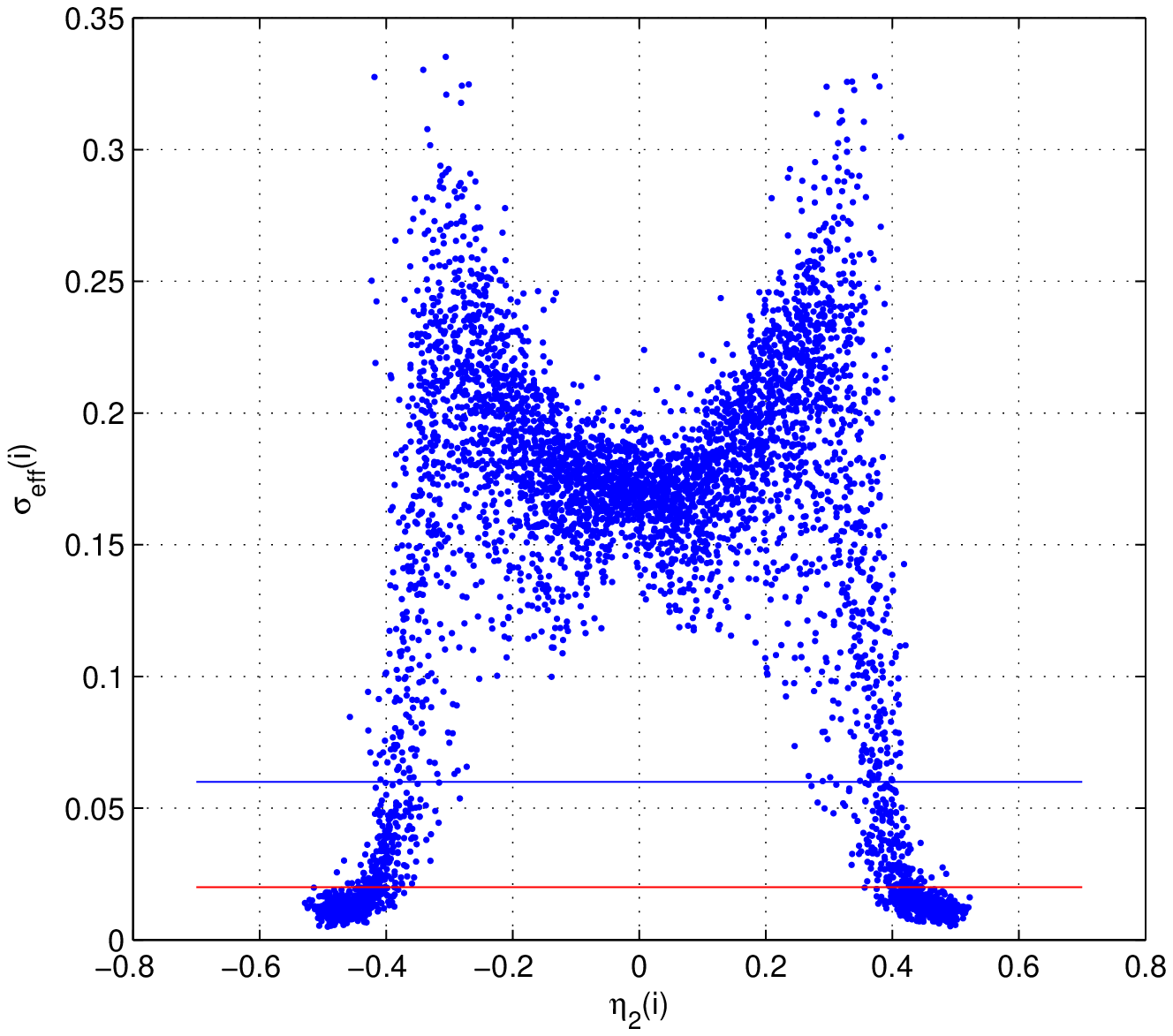}
\caption{\em Left figure. Scatter-plot of the $\sigma_{eff}(i)$ for the
floating strip side in function of the $\eta_2$ positions. Right figure. Scatter-plot of the  $\sigma_{eff}(i)$ for
the normal noisy side (with evident lower resolution). The horizontal lines will be illustrated
in the following.
}\label{fig:figure0a}
\end{center}
\end{figure}

The possibility of different statistical properties of each hit
is based on well known properties of the MIP hitting a silicon strip
(or gas discharge) detectors,
largely reported in literature:
a) the charge released in each hit has a Landau distribution with
different signal-to-noise ratio.
b) The position resolution depends from the charge spreading thus
it is better near to the strip border (e.g. ref.~\cite{CMSm} for gas ionization chambers)
c) Each strip has its own noise.
Thus the assumption of homoscedasticity (on an entire
detector layer) is inconsistent for the previous well known reasons,
which are common to a large class of detectors.
To handle
these random effects a complex mathematical infrastructure is required,
it must be able to generate different PDFs for each hit.
Few results of this mathematical infrastructure, illustrated in the follow, 
are reported in figure~\ref{fig:figure0a}. 
They are samples of the (calculated) effective variances for hits in the two 
sides of our micro-strip detectors.
Small effective variances at the strip borders can be observed in the
right part of figure~\ref{fig:figure0a} for the noisy normal strip side.
They share some similarity with the results of ref.~\cite{CMSm} supporting a
general aspect of this border effect. The floating strip side is almost always
much better, the vertical scale of the right plot is around three times that of the left plot.
%

%
%
The demonstration of different PDF for groups of hits, or as in our
case a different PDF for each hit, poses serious problems to the
use the standard least squares as a fitting tools. A frequently used
assumption to derive the least squares (or linear regression
as it is often called) properties is the identity of the variance of
the (hit) measurement errors (homoscedasticity). In this case,
it is easy to show the optimality and
the unbiasedness of the method. If the homoscedasticity breaks down, the
optimality and unbiasedness disappear. Linearity is the
only surviving feature. It must be clear that the loss of the optimality means a loss
in resolution, and the resolution is never too much. Similar
warnings must be raised for the detector architectures
simulated (and realized) starting from non-optimal methods. It is unclear the
origin of this trivialization of the Gauss approach~\cite{gauss} where heteroscedasticity
is assumed from the very beginning.

\subsection{A short derivation of the PDF for the COG$_2$ algorithm}

The incidence angle of our MIPs imposes the use of the minimum
number of signal strips to reduce the noise, hence, as stated above,
only two signal strips will be used.
In ref.~\cite{landi05} we indicated the principal steps required
to obtain the PDF for the COG$_2$ (two-strip COG), those steps followed
the standard method described in the books about the theory of probability.
At first one has to obtain the cumulative distribution with integrations on complex
surfaces, after, the derivative of the cumulative distribution gives the PDF.
The length and the complexity of this development (our first procedure) is too boring
to be published. A simpler and more
direct method applied to few COG$_n$ algorithms will be published elsewhere.
A small portion of that method is reported here.
In few steps, results identical to those
given by the cumulative distribution are obtained.
Here, the COG$_2$ algorithm has the origin of its reference system
in the center of the maximum-signal strip. The strip signals are indicated with:
$x_1$, $x_2$, and $x_3$, respectively the signal of the right strip,
central strip (with the maximum signal) and left strip. If $x_1>x_3$
the COG$_2$ is $x=x_1/(x_1+x_2)$, if $x_3>x_1$ it is $x=-x_3/(x_3+x_2)$. Thus:
\begin{equation}\label{eq:equation_1}
\begin{aligned}
    &P_{x_{g2}}(x)=\int_{-\infty}^{+\infty}\mathrm{d}\,x_1
    \mathrm{d}\,x_2
    \mathrm{d}\,x_3\,P(x_1,x_2,x_3)\\
    &\,\Big[\theta(x_1-x_3)\delta(x-\frac{x_1}{x_1+x_2})+
    \theta(x_3-x_1)\delta(x+\frac{x_3}{x_3+x_2})\Big]
\end{aligned}
\end{equation}
where $P(x_1,x_2,x_3)$ is the PDF to have the signals $x_1,x_2,x_3$
from the strips $1,2,3$. The signals $x_i$ are at their final elaboration stage
(pedestals, common noise, etc.) and ready to be used for the hit-position
reconstruction. The function $\theta(x_j)$ is the Heaviside $\theta$-function:
$\theta(x_j)=1$ for $x_j>0$, $\theta(x_j)=0$ for $x_j\leq 0$, and $\delta(x)$ is
the Dirac $\delta$-function.
The normalization of $P_{x_{g2}}(x)$ can be easily verified by an integration on $x$.
Equation~\ref{eq:equation_1} can be further elaborated with the following
transformations. Splitting the sum of eq.~\ref{eq:equation_1} in two independent
integrals and transforming the variables $x_1=\xi$, $x_1+x_2=z_1$ and $x_3=\beta$,
$x_3+x_2=z_2$, the jacobian of the transformation is one and the integrals
in $z_1$ and $z_2$ can be performed with the rule:
\begin{equation}\label{eq:equation_1a}
    \int_{-\infty}^{+\infty}\mathrm{d}\,z\, F(z-\nu)\,
    \delta(x\mp\frac{\nu}{z})=F(\frac{\pm\nu}{x}-\nu)\,\frac{|\nu|}{x^2}\, .
\end{equation}
Applying eq.~\ref{eq:equation_1a} to eq.~\ref{eq:equation_1} and using the limitations of the two
$\theta$-functions, eq.~\ref{eq:equation_1} becomes:
\begin{equation}\label{eq:equation_1b}
\begin{aligned}
    P_{x_{g2}}(x)=\frac{1}{x^2}\Big[&\int_{-\infty}^{+
    \infty}\mathrm{d}\,\xi\int_{-\infty}^\xi\,\mathrm{d}\beta\,
    P\big(\xi,\,\xi\frac{1-x}{x},\beta\big)\,|\xi|\,+\\
    &\int_{-\infty}^{+\infty}\mathrm{d}\,\beta\,\int_{-\infty}^\beta\,
    \mathrm{d}\xi\,P\big(\xi,\,
    \beta\frac{-1-x}{x},\beta\big)\,|\beta|\Big]\,.
\end{aligned}
\end{equation}
This form underlines very well the similarity with the Cauchy PDF; in the
limit of $x\rightarrow\infty$ the $x$-part of the $P(.)$ arguments are
$-1$ and $P_{x_{g2}}(x)\propto 1/x^2$ for large $x$.

\subsection{The probability $P_{x_{g2}}(x)$ for small $x$}

The probability $P(x_1,x_2,x_3)$ can handle a strict correlation
among its arguments, we release this strict correlation with the
weakest one: the mean values of the strip signals $\{a_i\}$ are
correlated, but the fluctuations around $\{a_i\}$ are independent.
Thus the probability $P(x_1,x_2,x_3)$ becomes the product of three
functions $\{P_i(x_i),i=1,2,3\}$. Each $P_i(x_i)$ is assumed to be
a gaussian PDF with mean values $a_i$ and standard deviation $\sigma_i$
(the strip additive noise is well reproduced by a gaussian in the absence
of MIP signals).
To simplify, the constants $a_i,i=1,2,3$ are the noiseless signals
released by a MIP with impact point $\varepsilon$:
\begin{equation}\label{eq:equation_1d}
    P_i(x_i)=\exp\big[-\frac{(x_i-a_i)^2}{2\sigma_i^2}
    \big]\frac{1}{\sqrt{2\pi}\sigma_i}.
\end{equation}
Even with the gaussian functions, the integrals of eq.~\ref{eq:equation_1b}
have no analytical expressions and effective
approximations must be constructed. We will not report our
final forms that are very long, instead we will illustrate a limiting
case which gives a simple approximation and eliminates a disturbing
singularity for the numerical integrations.
It easy to show that for $|x|\rightarrow 0$, $x^{-1}P_2\big(\xi(1-x)/x\big)$ and
$x^{-1}P_2\big(\beta(-1-x)/x\big)$ converge to two Dirac $\delta$-functions.
Hence, for small $|x|$ (or better for $|x|\sigma_2/a_2\ll1$),
the integrals of eq.~\ref{eq:equation_1b} can be expressed as:

\begin{equation}\label{eq:equation_1c}
\begin{aligned}
    P_{x_{g2}}(x)=\frac{|a_2|}{\sqrt{2\pi}}\Big\{&
    \frac{\exp\big[-(x-\frac{a_1}{a_1+a_2})^2
    \frac{(a_1+a_2)^2}{2\sigma_1^2(1-x)^2}\big]
    \big[1-\mathrm{erf}\big((\frac{a_3}{a_2+a_3}-x)
    \frac{a_2+a_3}{\sqrt{2}\sigma_3|1-x|}
    \big)\big]}{2\sigma_1(1-x)^2}+\\
    &\frac{\exp\big[-(x+\frac{a_3}{a_3+a_2})^2
    \frac{(a_3+a_2)^2}{2\sigma_3^2(1+x)^2}\big]
    \big[1-\mathrm{erf}\big((\frac{a_1}{a_2+a_1}+x)
    \frac{a_2+a_1}{\sqrt{2}\sigma_1|1+x|}
    \big)\big]}{2\sigma_3(1+x)^2}\Big\}\,.
\end{aligned}
\end{equation}

Equation~\ref{eq:equation_1c} is correct for $|x|\rightarrow 0$, but it is
useful beyond this limit, as far as the pole for $x=\pm1$
is irrelevant, and contains many ingredients of more complex expressions.

\begin{figure}[ht]
\begin{center}
\includegraphics[scale=0.6]{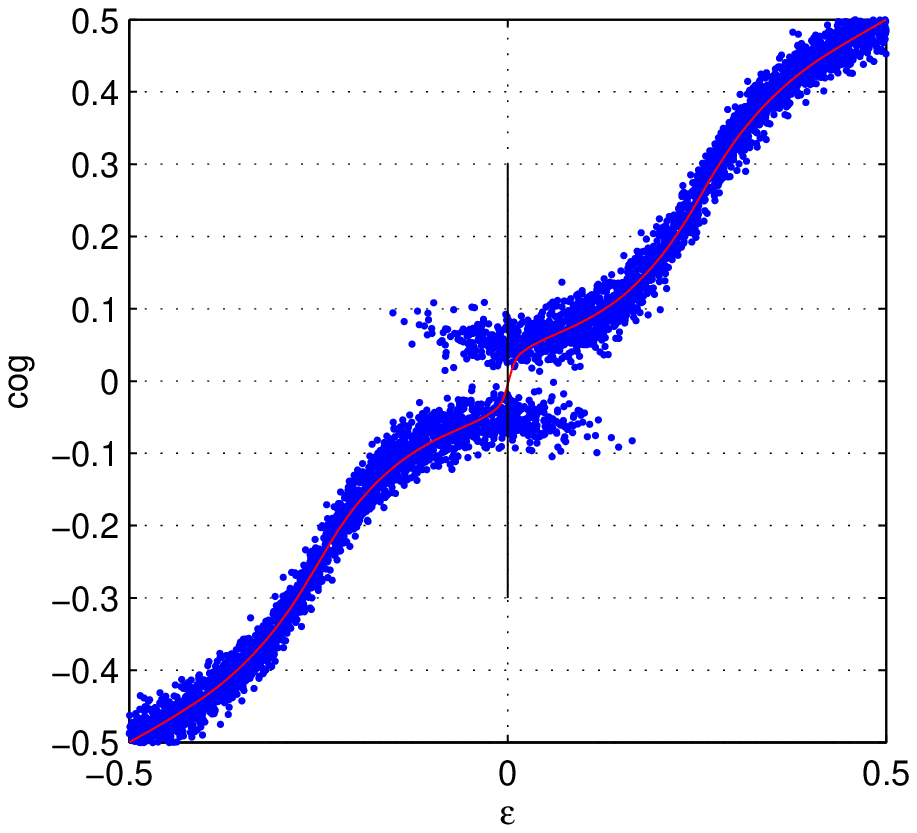}
\includegraphics[scale=0.7]{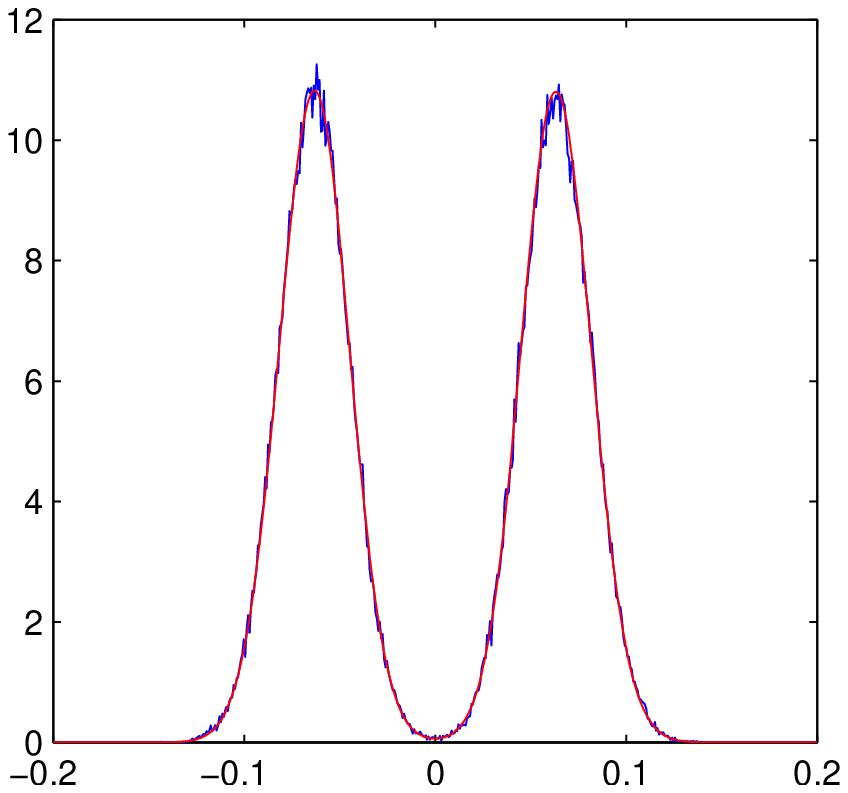}
\caption{\em Left plot. Blue dots: distribution of COG$_2$
for the floating strip side, continuous line $\eta_2$ algorithm.
Right plot: red line, PDF of eq.2.5
along the black line of the left-side figure. Blue line: the simulated COG$_2$
PDF with  Landau fluctuations along the track (200 K-events)
}\label{fig:figure0}
\end{center}
\end{figure}

An example of eq.~\ref{eq:equation_1c} can be seen in the right
side of figure~\ref{fig:figure0}.
The essential elements
are the two maxima (gaussian-like)  centered in the possible
noiseless two-strip COG: $a_1/(a_1+a_2)$ and $-a_3/(a_3+a_2)$.
The widths of the two maxima are modulated by the signal-to-noise
ratio $(a_2+a_1)/\sigma_1$ and $(a_2+a_3)/\sigma_3$, the $1\pm x$
factor is relevant at the borders of the  $x$-range, i.e.
$x=\pm 0.5$. Each maximum has a scaling factor $|a_2|/\sigma_1$ or
$|a_2|/\sigma_3$ that, for the large majority of strip clusters
$\sigma_1\approx\sigma_2\approx\sigma_3$, is
the signal to noise ratio of the central strip.

Increasing the impact position $\varepsilon$,
the lowest maximum tend to disappears. Similarly at decreasing $\varepsilon$,
the highest maximum is rapidly reduced. The tails of the maxima differ
drastically from gaussian PDF.
More complete expressions of eq.~\ref{eq:equation_1c} contain
terms very similar to Cauchy PDFs, they have no $a_j$ factors
and survive when all the $a_j$ are zero.

The dimensions of the constants $a_j$ must be those of the $\sigma_j$,
for both of them we take directly the ADC counts. The $x$-variable
(the COG$_2$) is a pure number expressed as a fraction of the strip
size, or more precisely, the strip size is the scale of lengths. In
the simulated distribution of figure~\ref{fig:figure0}, the particle
track is split in 15 segment and each segment has a different ionization
density, obtained from a Landau PDF accorded to the length of the segment.
The total ionization is kept constant. The ionization of each segment is independently
diffused toward the collecting strips and scaled to the signal mean value,
a gaussian noise of 4 ADC counts is added to each strip. The distribution
of the COG$_2$ is reported in the normalized histogram. At this orthogonal
incidence the Landau fluctuations are almost entirely contained in the
fluctuations of the total collected charge.

\subsection{The functional dependence from the impact point}

For our fitting task, the PDFs with constant $a_i$, the
noiseless version of the strip signals, and variable $x$ (the COG$_2$ noisy values),
are irrelevant. We need the PDFs of the impact point $\varepsilon$ at constant
(noisy) $x$. It is clear that the mean values of eq.~\ref{eq:equation_1d} depend
from the impact points and they are related to $x$ by eq.~\ref{eq:equation_1c}.
To simplify this dependence we redefine as $a_i(\varepsilon)$ the fraction of
signal acquired by each strip in function of $\varepsilon$, an additional parameter
to normalize this fraction must be recovered from data of each hit.
The constructions of the functions $a_i(\varepsilon)$ has been
illustrated in ref.~\cite{landi05}, for orthogonal incidence. Identical
expressions will be used here. These new pieces of
information, contained in the $\{a_i(\varepsilon)\}$,
extend the PDF of eq.~\ref{eq:equation_1b} to be
function of the impact point. The PDF can be rewritten as:

\begin{equation}\label{eq:equation_2}
    P_{x_{g2}}(x,E_t,\varepsilon)=\frac{F(a_1(\varepsilon),
    a_2(\varepsilon),a_3(\varepsilon),E_t,\sigma_1,\sigma_2,\sigma_3,x)}{x^2}\,.
\end{equation}
Where $F(a_1(\varepsilon),a_2(\varepsilon),a_3(\varepsilon),E_t,\sigma_1,\sigma_2,\sigma_3,x)$ is
the term in square brackets
of eq.~\ref{eq:equation_1b}. Equation~\ref{eq:equation_2} is the mathematical
infrastructure able to connect the points a), b), c) discussed above.
The functions $a_1(\varepsilon),a_2(\varepsilon),a_3(\varepsilon)$
are the fractions of signal collected by the strips $1,2,3$ in function of $\varepsilon$,
they are normalized with $E_t$,  the total signal
of the three strips: the central one with the maximum signal and the two lateral.
The extraction of the functions $\{a_i(\varepsilon)\}$ from
real data described in ref.~\cite{landi05} is a delicate operation, but a strict compliance to the
described equations produces very reliable results.
In any case, slight variations of the $\{a_i(\varepsilon)\}$ around the best one give almost
identical track parameter distributions, thus their selection is very important but less critical than expected.
Our function $a_2(\varepsilon)$ has a definition similar to the functions called "templates" in ref.~\cite{CMS}.
The parameters $\sigma_1,\sigma_2,\sigma_3$
are the standard deviations (eq.~\ref{eq:equation_1d}) of the three strips considered.
Equation~\ref{eq:equation_2} can easily handle strips with different $\sigma$, but,
in the simulations, we will use identical $\sigma$s for all the strips of same detector side.
To easy the notations, these parameters will not be
reported in the future expressions. Equation~\ref{eq:equation_2} is normalized as function of $x$ but
is not normalized in $\varepsilon$. This normalization is irrelevant for the maximum
likelihood search and will be neglected here.
A set of normalized PDFs from eq.~\ref{eq:equation_2} are illustrated in ref.~\cite{landi05} as $\varepsilon$-functions.
It should be evident the enormous number of different PDFs that can be generated by eq.~\ref{eq:equation_2},
it depends from five parameters and three independent functions (essentially
at least hundred free parameters). For the COG$_3$ case, the equivalent of
eq.~\ref{eq:equation_2} depends from seven parameters and five independent functions, at large incidence
angle this type of PDF is one of the candidates for the fit.

\subsection{The $\eta$ Position Algorithms}

Two different position algorithms will be used in the following: the COG$_2$
algorithm and the $\eta_2$ algorithm. The results of
the least squares are better with the $\eta_2$ positioning algorithm than with the COG$_2$.
The $\eta_2$-algorithm is built to correct the COG$_2$ systematic error.
Assuming a definite relation of the impact point $\varepsilon$ with the
COG$_n$, the PDF of the COG$_n$ can be defined as the coordinate transformation:

\begin{equation}\label{eq:equation_2a}
    p(\varepsilon)\Big|\frac{d\varepsilon}{d x_{gn}}\Big|=\Gamma_n(x_{gn})\,,
\end{equation}
where $\Gamma_n(x_{gn})$ is the PDF of COG$_n$  given by $p(\varepsilon)$ as PDF of
impact points. For uniform distribution of the impact point $p(\varepsilon)=1/\tau$
and normalized to one on the strip length $\tau$ ($\tau=1$ here), the
differential equation can be integrated:

\begin{equation}\label{eq:equation_2b}
    \varepsilon_n(x_{gn})=\varepsilon(x_{gn}^0)+\int_{x_{gn}^0}^{x_{gn}}\Gamma_n(y)\, dy\,,
\end{equation}
the positive sign of $d\varepsilon/d x_{gn}$ is discussed in refs.~\cite{landi01,landi03}.
With the exact initial constant $\varepsilon(x_{gn}^0)$, the function
$\varepsilon_n(x_{gn})$ is a better estimation of the impact
point than the simpler COG$_n$.  In the $\eta_2$
positioning algorithm, the function $\varepsilon_2(x_{g2})$ is
given by the experimental COG$_2$ PDF inserted  in eq.~\ref{eq:equation_2b}.
The extraction of the initial constant $\varepsilon(x_{g2}^0)$ from the data is
discussed in refs.~\cite{landi03,landi04} and tested in a dedicated test
beam~\cite{vannu}. The definition of eq.~\ref{eq:equation_2a} generalizes
the $\eta$ algorithm of ref.~\cite{belau} for any COG$_n$. Our $\eta_2$
algorithm coincides with that of ref.~\cite{belau} with $\varepsilon(x_{g2}^0)= x_{g2}^0=0$.
This condition is true for an exact axial symmetry of the signal
distribution and of the strip response function. It is evident that real detectors
drastically deviate from this symmetry, and essential corrections must be
implemented for $\varepsilon(x_{g2}^0)$.

\subsection{Track definition}

Given the novelty of this approach, our track definition must be very simple.
The tracks are circles with a large radius to simulate the high momentum MIPs
where the multiple scattering is negligible. The relation of the track parameters
to the momenta is $p=0.3\, \mathrm{B}\, R$ (ref.~\cite{particle_group}) here $p$ is in $\mathrm{GeV/c}$,
$\mathrm{B}$ in Tesla and $R$, the track radius, in meters.\newline
The tracker model is formed by six parallel equidistant ($89\, \mathrm{mm}$)
detector layers as in the PAMELA tracker. The constant
magnetic field is $0.44\,T$. The simulated tracks are on
a plane perpendicular to the the detector layers and to the
magnetic field, ($\xi\,,z$) are the coordinates of the
track plane. The center of the tracker is at
$\xi=0\,,z=0$, the $z$ axis is parallel to the
layer planes and the $\xi$ axis is perpendicular.
The tracks are circles with center in $\xi=-R$ and $z=0$,
and the magnetic field is parallel to the analyzing strips.
To simplify the geometry, the overall small rotation of the Lorentz angle ($0.7^\circ$)  is neglected now, but it will
be introduced in the following.
The simulated hits are generated (ref.~\cite{landi05}) with a uniform random distribution on a strip,
they are collected in groups of six to produce the tracks. The exact impact position $\varepsilon$ of each hit is subtracted from
its reconstructed $\eta_2(x_{g2})$ position (as defined in refs.~\cite{landi05,landi03,belau}) and it is added the value of the
fiducial track for the corresponding detector layer. In this way each group of six hits
defines a track with our geometry and the error distribution of $\eta_2$-algorithm. Identically for the
COG$_2$ positioning algorithm.
This hit collection
simulates a set of tracks populating a large portion
of the tracker system with slightly non parallel strips on different layers (as it is always  in real detectors).
At our high momenta the track bending is very small and the track sagitta is smaller than the strip width.
Thus, the bunch of tracks has a transversal section around a strip size, on this width we have to
consider the Lorentz angle but its effect is clearly negligible.
Our preferred position reconstruction is the $\eta_2$ algorithm as in ref.~\cite{landi05}  because it
gives parameter distributions better than those obtained with the simplest COG$_2$
positions, but even the results for the COG$_2$ will be reported in the following.\newline
In the $\{{\xi},{z}\}$-plane, the circular tracks are approximated, as usual,  with parabolas linear in the
track parameters:
\begin{equation}\label{eq:equation_3}
\begin{aligned}
    &\xi=-R+\sqrt{R^2-{z}^2}\approx \beta+\gamma-\alpha z^2=\varphi(z)\,\\
    &\xi_n=\beta_n+\gamma_n {z}-\alpha_n {z}^2=\varphi_n(z)\,.\\
\end{aligned}
\end{equation}
The first line of eq.~\ref{eq:equation_3} is
the model track, the second line is the fit result.
The circular track is the osculating circle of the parabola
$\varphi(z)$, at our high momenta and tracker size
the differences are negligible. The function $\varphi(z)$ has
$\gamma=0,\,\beta=0$, and $1/\alpha$ is proportional
to the track momentum. Due to the noise, the
reconstructed track has equation $\varphi_n(z)$ and
the parameters $\{\alpha_n,\beta_n,\gamma_n\}$, given
by the fit, are distributed around the model values.
The non gaussian forms of our PDFs oblige the use of
the non-linear search of the likelihood maxima,
and, as always, the search will be transformed in a
minimization. The momentum and the other parameters
of the track $n$ are obtained minimizing, respect to
$\{\alpha_n,\beta_n,\gamma_n\}$, the function
$L(\alpha_n,\beta_n,\gamma_n)$ defined as the
negative logarithm of the likelihood with the
PDFs of eq.~\ref{eq:equation_2}:
\begin{equation}\label{eq:equation_4}
\begin{aligned}
    &L(\alpha_n,\beta_n,\gamma_n)=-\sum_{j=6n+1}^{6n+6}\,
    \ln[P_{x_{g2}}(x(j),E_t(j),\psi_j(\alpha_n, \beta_n,\gamma_n)]\\
    &\psi_j(\alpha_n,\beta_n,\gamma_n)=\varepsilon(j)-\varphi(z_j)+
    \varphi_n(z_j)\,.
\end{aligned}
\end{equation}
The parameters $x(j)$, $E_t(j)$ (introduced in
eq.~\ref{eq:equation_2}) are respectively:
the COG$_2(j)$ position, the sum of signal in the
three strips for the hit $j$ of the track $n$. The term
${z}_j$ is the position of the 
detector plane $j$ of the $n$th track.
The $\varepsilon$-dependence in the $\{a_i(\varepsilon)\}$
is modified to $\psi_j(\alpha_n,\beta_n,\gamma_n)$,
to place the impact points on the track. In real data,
$\varepsilon(j)-\varphi(z_j)$ is
absent (and unknown) but the data are supposed to be on a
track. We can easily use  a non linear form for the
function $\psi_j(\alpha_n,\beta_n,\gamma_n)$, but in this
case is of scarce meaning. In more complex cases,
non linearities of various origin can be easily implemented, for example
analytic expressions of the tracks that exactly account
for the magnetic field inhomogeneities.

We will reserve the definition of Maximum Likelihood Evaluation
(MLE) to the results of eq.~\ref{eq:equation_4}.
The routine for the MLE  is initialized as in
ref.~\cite{landi05}, the first track parameters are
given by a weighted least squares with weights given by an
effective variance ($\sigma_{eff}(i)^2$)
for each hit $(i)$ and $\eta_2(i)$ as hit position. The
$\sigma_{eff}(i)^2$ is obtained from eq.~\ref{eq:equation_2}, but,
for its form, the variance is an ill defined
parameter even in $\varepsilon$. Cuts in the integration
limits suppress the tails and give finite results.

\begin{equation}\label{eq:equation_4a}
    \sigma_{eff}(i)^2=\int_{\eta_2(i)-c_t}^{\eta_2(i)+c_t}
    P_{x_{g2}}(x(i),E_t(i),\varepsilon)\varepsilon^2\,\mathrm{d}\varepsilon\,.
\end{equation}
The parameter $c_t$  is selected to reproduce the PDFs $P_{x_{g2}}(x(i),E_t(i),\varepsilon)$
with gaussian functions of variance $\sigma_{eff}(i)^2$ in the case of excellent hits.
For a set of hits (excellent hits), eq.~\ref{eq:equation_2}
has the form of a narrow high peak and a gaussian, centered in
$\eta_2(i)$ and variance $\sigma_{eff}(i)^2$, can overlap well the
$P_{x_{g2}}(x(i),E_t(i),\varepsilon)$ around the maximum.
We selected few of them for the tuning of $c_t$. These gaussian
approximations look good in linear plots, the logarithmic plots show
marked differences even in these happy cases: the tails are non-gaussian.
The cuts, so defined, are used for the $\sigma_{eff}(i)$ extraction for
all the other hits, even where $P_{x_{g2}}(x(i),E_t(i),\varepsilon)$ is
poorly reproduced by a gaussian. We use two sizes of cuts, one for each side
of our detector.

The $\{\alpha_n,\beta_n,\gamma_n\}$ given by the weighted least
squares are almost always near those given by the minimization
of eq.~\ref{eq:equation_4}, thus accelerating the
minimum search to the MATLAB~\cite{MATLAB} {$fminsearch$} routine.
The closeness of these approximations to the MLE supports the
non criticality of the extraction of the functions $\{a_i(\varepsilon)\}$.
The approximate gaussian distributions
are often very different from the hit PDFs
but this partial information suffices
to produce near optimal results.
When the tails of the PDFs are
important the MLE gives better results.
The strong variations of the set $\{\sigma_{eff}\}$ on the
strip are illustrated in figure~\ref{fig:figure0a}.
The time-consuming extraction of $\sigma_{eff}(i)$
can be reduced building a look-up table of $\sigma_{eff}(\eta_2(x),E_t)$
and calculating $\sigma_{eff}(i)$ with an interpolation.
This use of eq.~\ref{eq:equation_4a}
to get weights to insert in the least squares is directly consistent with ref~\cite{gauss}.

Having to plot the results of four type of reconstructions we will use the following color
convention for the lines:
\begin{itemize}
  \item {\bf{red lines}} refer to our MLE  (eq.~\ref{eq:equation_4}),
  \item {\bf black lines} are the weighted least squares with weight $1/\sigma_{eff}(i)^{2}$ and $\eta_2$ position,
  \item {\bf blue lines} for the least squares with the $\eta_2$ position algorithm
  \item {\bf magenta lines} for the least squares with the COG$_2$ position algorithm.
\end{itemize}

\begin{figure}[ht]
\begin{center}
\includegraphics[scale=0.45]{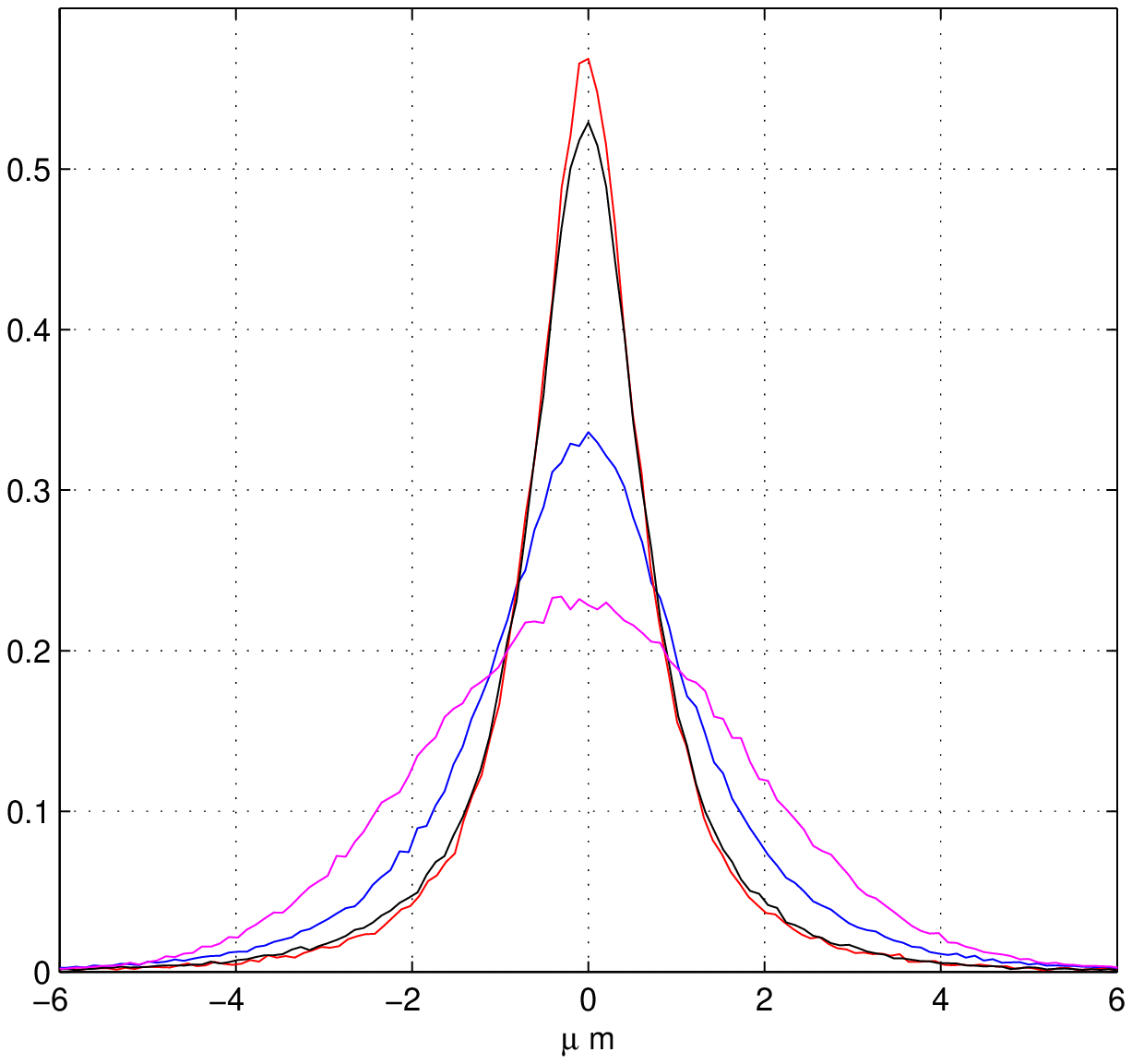}
\includegraphics[scale=0.45]{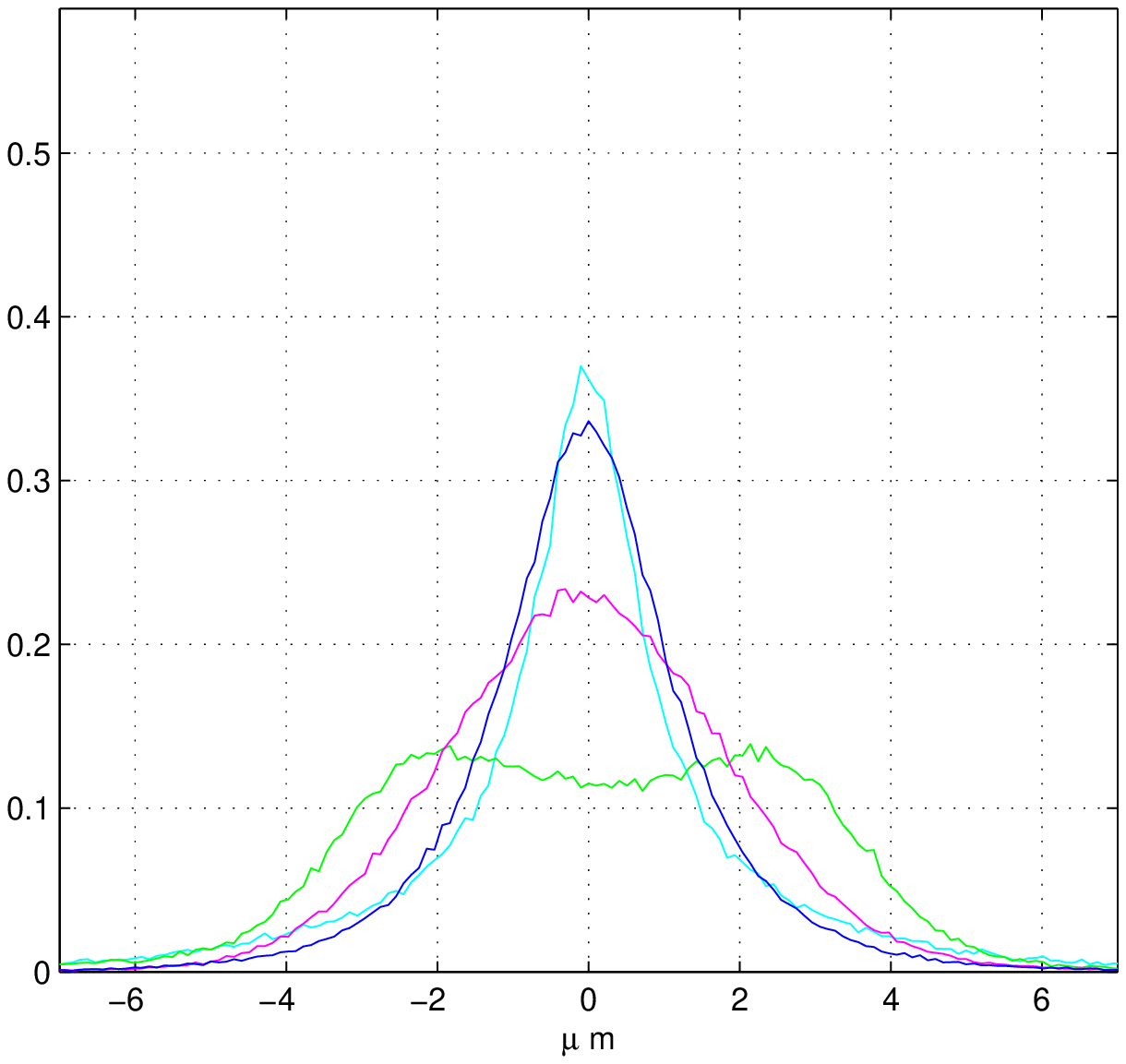}
\caption{\em Left plot. True residuals of the
reconstructed tracks.  MLE (red), schematic model $\sigma_{eff}(i)$
(black), hit position $\eta_2$ (blue), hit position COG$_2$ (magenta).
Right plot,  position errors of the $\eta_2$ (cyan)
true residuals for the hit position $\eta_2$ (blue), true residual for hit position COG$_2$ (magenta),
and errors of the COG$_2$ (green).
}\label{fig:figure01}
\end{center}
\end{figure}

\section{Low noise, high resolution, floating strip side}

The floating strip side is the best of the two sides of this type of
strip detector. It is just this side that measures the track bending
in the PAMELA magnetic spectrometer.
In the test beam~\cite{vannu}, the noise PDF of the "average" strips
without signals is well reproduced by a gaussian with a
standard deviation of 4 ADC counts. The PDF for the sum of
three strip signals has its maximum at 142 ADC counts with
the most probable signal-to-noise ratio ($SNR$) of $35.5$
for a three strip cluster  ($SNR(n)=\sum_{i=1}^3 x_i(n)/\sigma_i$).
The functions $a_j(\varepsilon)$ are those of ref.~\cite{landi05} for this strip type.
In the simulations we will use a high momentum of $350 \, GeV/c$.
For this momentum and identical geometry, we have a report with some
histograms of a CERN test beam of the PAMELA tracker before its installation
in the satellite. The tracks were reconstructed with the original $\eta$ algorithm
of ref.~\cite{belau}, the curvature histogram turns out very similar to our, giving
an excellent check of our simulations. In this case with the orthogonal incidence
of the beam, the systematic error of $\eta$ algorithm of ref.~\cite{belau}
is constant (and small) and has no effect on the curvature reconstruction.
In the left side of figure~\ref{fig:figure01},  the
histogram values divided by the number of entries and the step size (frequency polygons
called distributions in the following)
of the differences of the fitted positions respect to the exact ones
({\em true residuals} in the following)
are reported. We will use the definition of true residuals (allowed only in simulations) to
distinguish from the {\em residuals} that generally indicate the differences from the fitted
positions with respect to the reconstructed ones. The highest curve is
the distribution of the true residuals of  our MLE,
followed immediately by the true residuals of our schematic model. The true residuals of the
least squares with $\eta_2$ and COG$_2$ as hit positions are the lowest two distributions.
For the absence of the
COG systematic error, the PDFs of the track parameters
given by the $\eta_2$ algorithm are  better than those given by the COG$_2$.
The right side of figure~\ref{fig:figure01} illustrates the relations of the true residual PDFs
for the least squares with the $\eta_2$ and COG$_2$ as hit positions with their corresponding
error PDFs. We see that, contrary to the general expectancy, the data redundancy does not
improves the position reconstructions for the $\eta_2$ algorithm. This result is similar to
the use of the least squares with data extracted from a Cauchy distribution: the least squares
degrades the original position resolution.
Instead, the  true residual PDF of COG$_2$ least squares looks better than
the error PDF. The least squares method looks able
to round the error distribution, but this effort consumes all its power.
In fact if the statistical noise is suppressed, the COG
systematic error does not allow any modification of the true residual PDF.
On the contrary,  the true residual PDF of the $\eta_2$ least squares grows
toward a Dirac $\delta$-function in the absence of
the statistical noise.

The PDFs of the residuals (as defined above) are not reported.
Those for the COG$_2$ and $\eta_2$ are
almost identical to the true residual PDFs of figure~\ref{fig:figure01}.
The residuals for the MLE and of the schematic model are very different
from their true residuals of figure~\ref{fig:figure01}. Narrow and high
peaks centered in zero are present in those two distributions. The peak of
the schematic model is higher than that of the MLE, and both of them
are higher than their maxima of the true residual distributions.
Our PDF of eq.~\ref{eq:equation_2} and the weights $1/\sigma_{eff}(i)^2$
distinguish good hits from bad hits and the fit optimization forces the track to
pass near to good ones, giving to them an high frequency of small residuals.
Often the residuals are used as a measure of the detector resolution, it
is easy to show the inconsistency of this assumption almost
in any case. For example in the right side of figure~\ref{fig:figure01} the distribution
of the true residuals  of the COG$_2$ is very similar to the residual distribution, but
it is very different from the COG$_2$ error distribution. In the following other similar
discrepancies will be underlined.

\begin{figure}[ht]
\begin{center}
\includegraphics[scale=0.45]{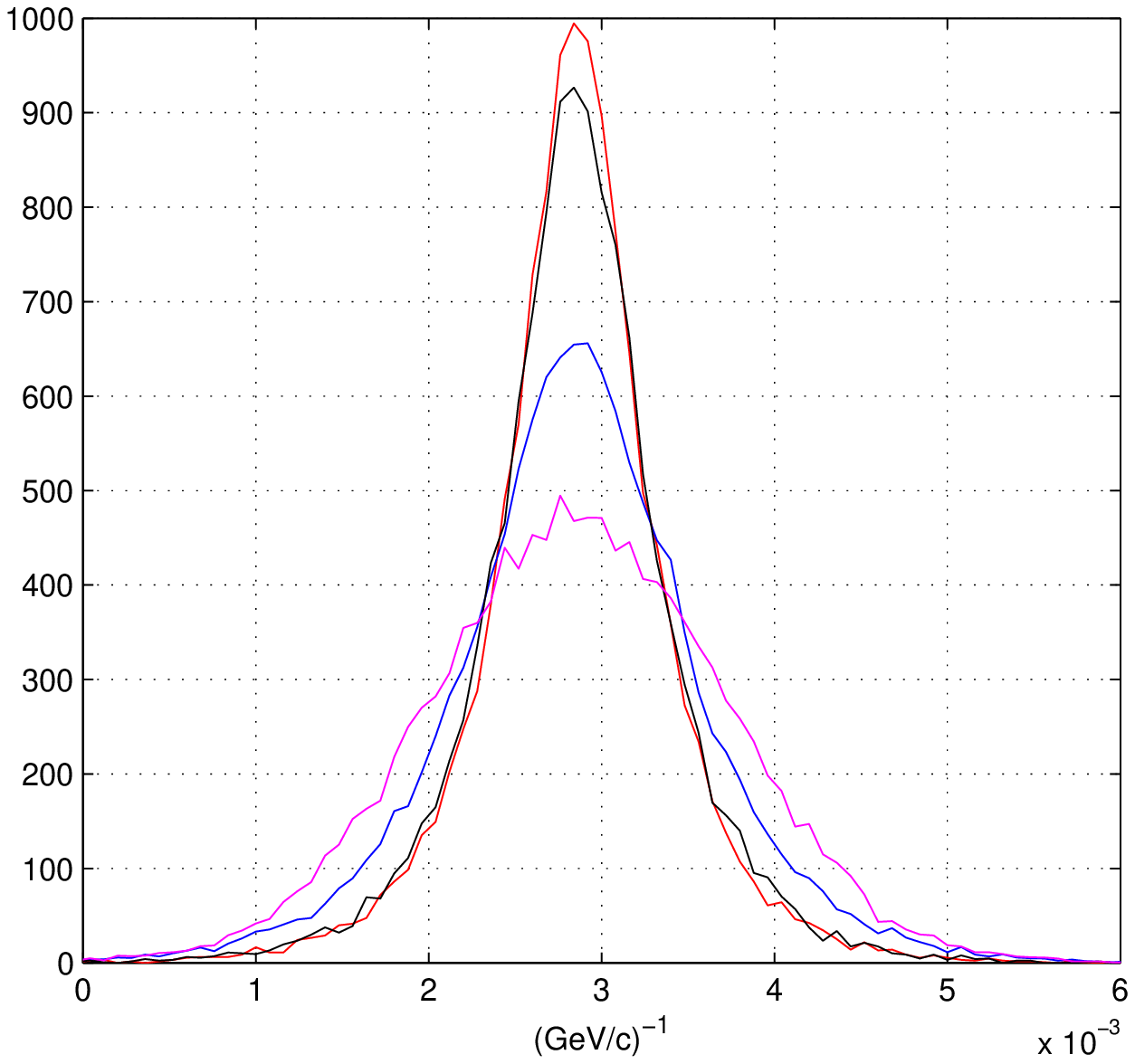}
\includegraphics[scale=0.45]{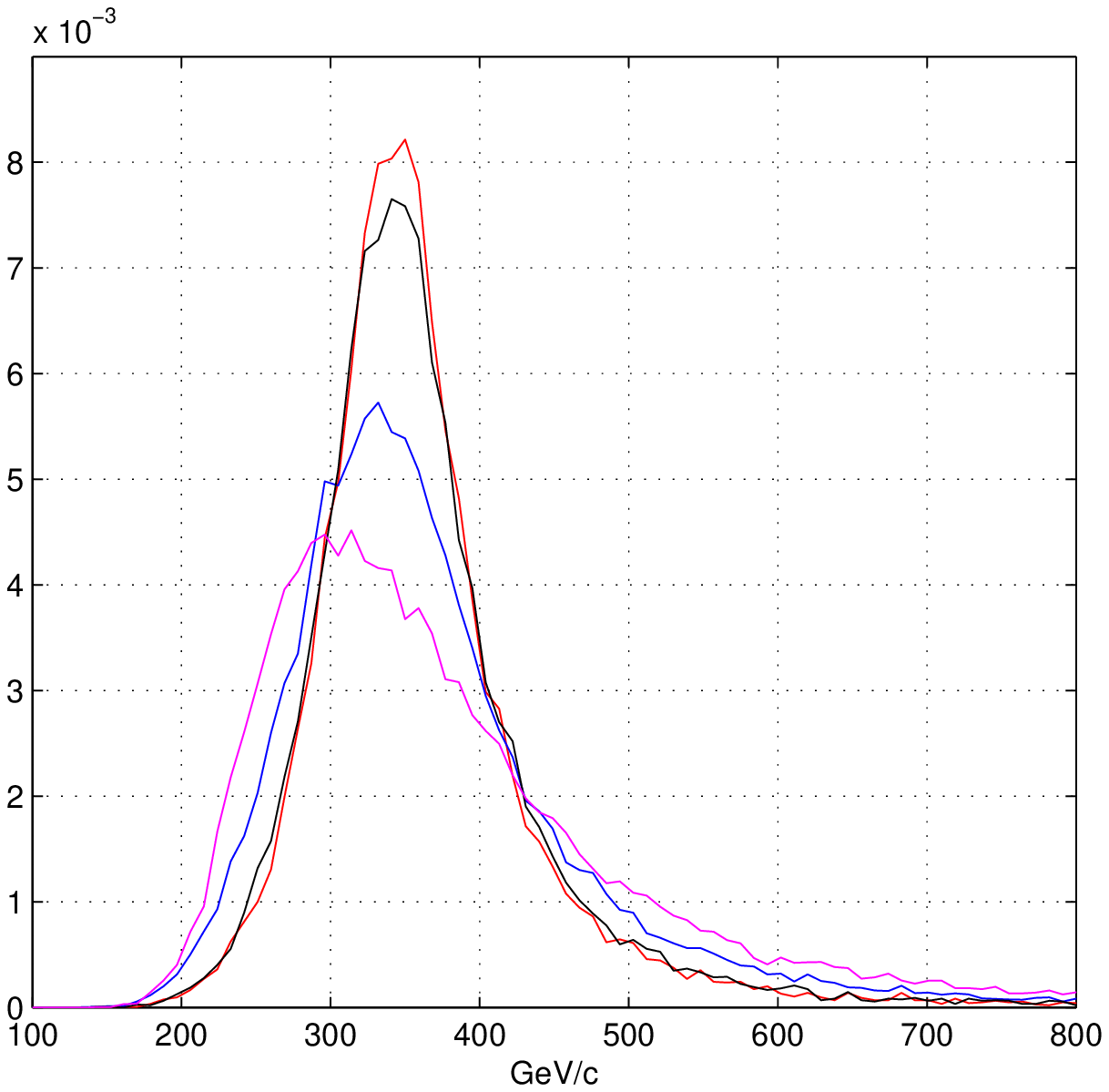}
\caption{\em Left plot. Distributions of the curvature in
$(GeV/c)^{-1}$ (the $\alpha$ parameters) MLE (red), schematic model $\sigma_{eff}(i)$ (black),
hit position $\eta_2$ (blue), hit position COG$_2$ (magenta).
Right plot:reconstructed momentum distributions. Color code as in the left plot.
}\label{fig:figure02}
\end{center}
\end{figure}
As illustrated in figure~\ref{fig:figure02}, the MLEs  give the
best results for the momentum reconstruction,
and the weighted least squares, are very near to them. The
fits of the standard least squares with $\eta_2$
or COG$_2$ positioning algorithms show a drastic decrease
in resolution. The use of the simple
COG$_2$ algorithm is the worst one.
Often the distributions of the left side of figure~\ref{fig:figure02}
are reported as resolution of the momentum reconstruction,
the k-value of ref.~\cite{particle_group}. For the $\eta_2$ and
COG$_2$ least squares, the plots of the momentum distributions
have appreciable shifts of the maxima (most probable value) respect
to the fiducial value of 350 $GeV/c$, the shifts are negligible
for the other two fits. These shifts are mathematical consequences
of the change of variable from curvature to momentum. For gaussian PDF, the shift
of the maximum due to the variable conversion from curvature ($GeV^{-1}$)
to momentum ($GeV$) is easily calculable, and depends
from the variance of the gaussian, increasing the variance the
maximum for the momentum PDF moves to lower values.

\subsection{Other track parameters}

The complete track reconstruction must consider even the other
two parameters of a track, the $\beta_n$ and $\gamma_n$. Their
fits give very similar distributions to those plotted in ref.~\cite{landi05}.
The maxima of the distributions are now a little lower, in
particular for the $\beta$ parameter. This is not unexpected,
in ref.~\cite{landi05} we had 3 degree of freedom for two
parameters, here we have 3 degree of freedom for three
parameters, an effective reduction  of the redundancy that
has a slight effect on the results.


\section{High noise, low resolution, normal strip side}
The other side of the double sided silicon microstrip detector
has very different properties respect to the floating strip
side (the junction side). We will not recall the special treatments
required to transform a ohmic side in strip detector  and all the
other particular setups necessary to its functioning. From the point
of view of their data, this side produces data very similar to
a normal strip with a gaussian noise of 8 ADC counts,
twice of the other side (most probable signal-to-noise ratio
$SNR(n)=18.2$). The absence of the floating strips gives to the
histograms of the COG$_2$ the normal aspect with a high central
density and a drop around $x_{g2}=0$.
No additional rises around $x_{g2}=\pm1/2$ are present, they are
typical of the charge spreads given by the floating strips.
This absence reduces the efficiency of the positioning algorithms
that gain substantially from the charge spread. Now, the functions
$\{a_i(\varepsilon)\}$ of ref.~\cite{landi05} are similar to those
of an interval function with a weak rounding to the borders,
this rounding is mainly due to the convolution of the strip response with
the charge spread produced by the drift in the collecting field.
If a residual capacitive coupling is present, it is very small. In
any case, it is just this rounding that renders very good the resolution
of the hits on the strip borders with a small $\sigma_{eff}(i)$ (e.g.
the points below the lowest horizontal line in figure~\ref{fig:figure0a}).
Due to the differences respect to the other side, and to obtain a
reasonable momentum distribution we have to implement
the simulations with a lower momentum of $150\, GeV/c$. Operating at
higher momentum the curvature distributions has a large part at negative
values and the momentum distribution acquires a second maximum at
wrong negative momenta.

\begin{figure}[ht]
\begin{center}
\includegraphics[scale=0.45]{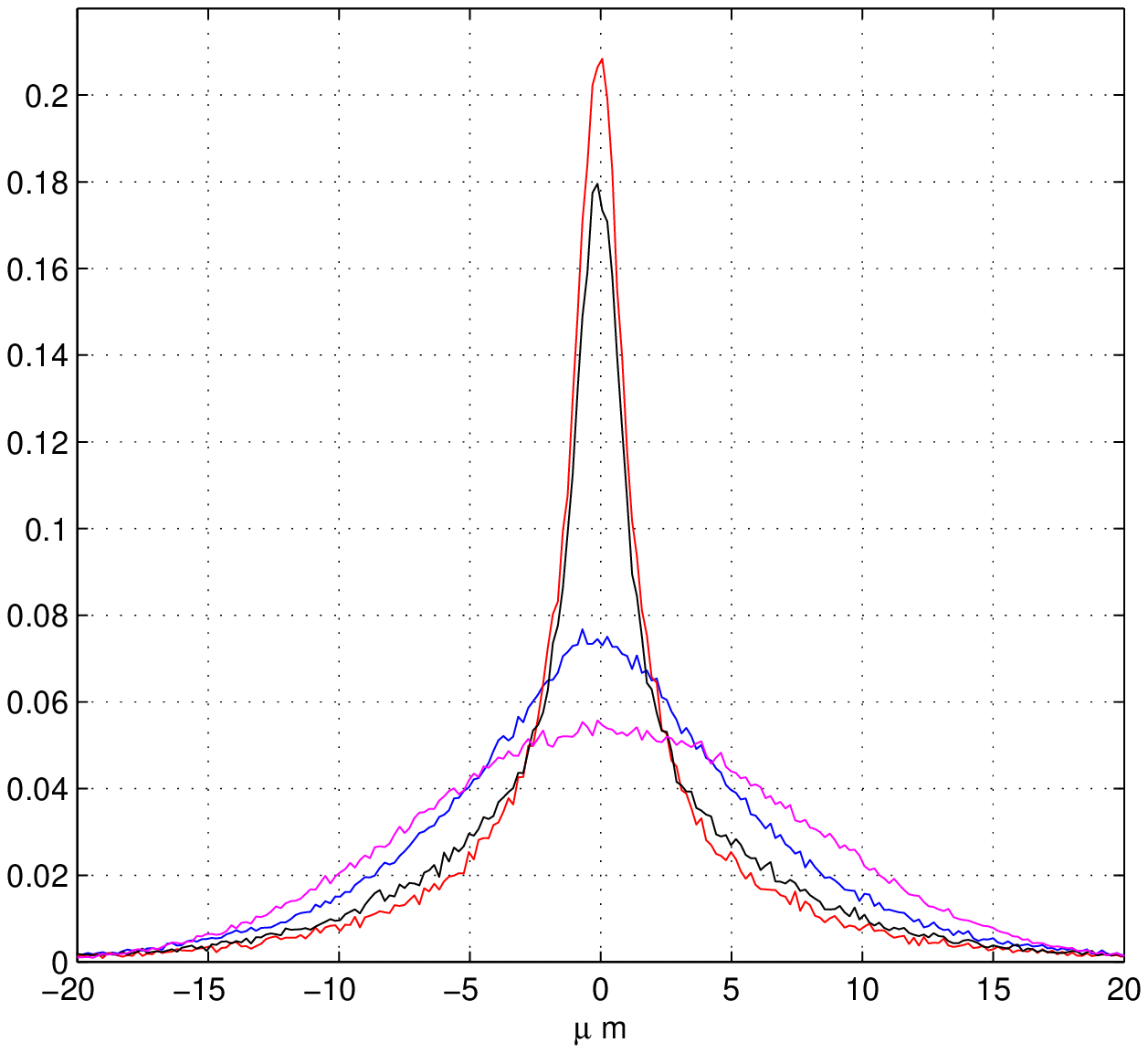}
\includegraphics[scale=0.45]{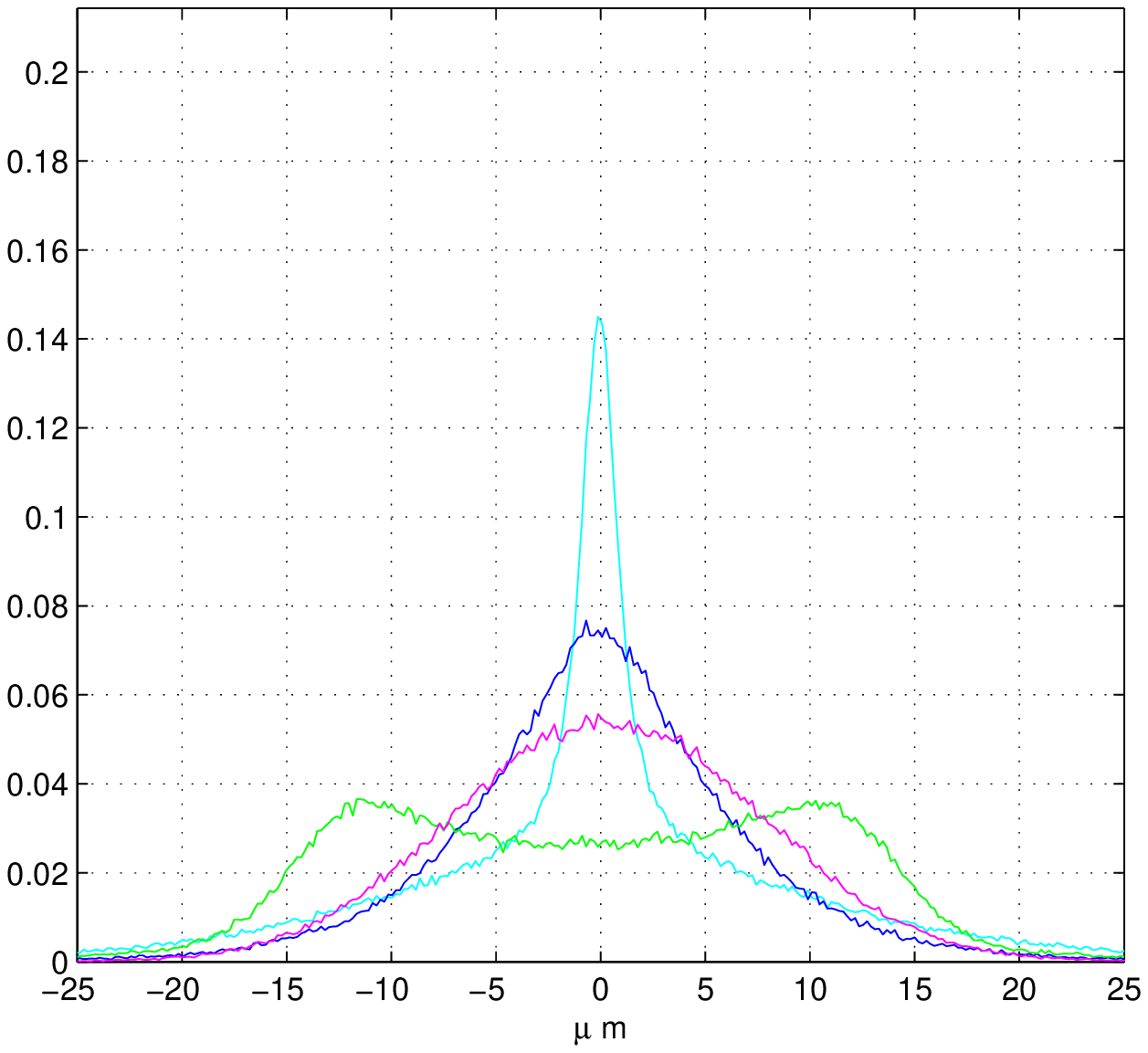}
\caption{\em The noisy normal strip side. Distributions of simulated results for
reconstructed tracks as defined in figure3.
}\label{fig:figure03}
\end{center}
\end{figure}

\begin{figure}[ht]
\begin{center}
\includegraphics[scale=0.45]{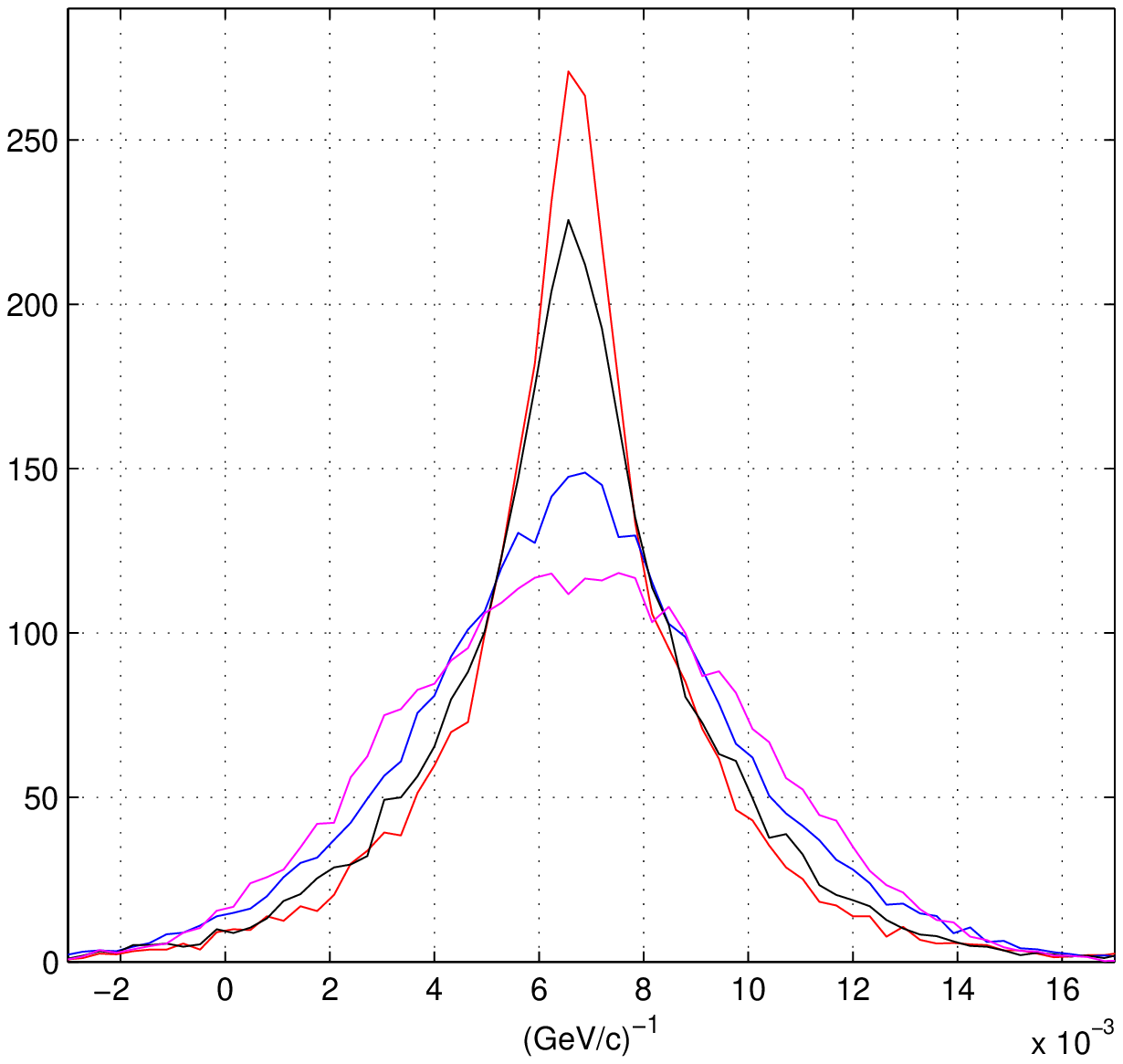}
\includegraphics[scale=0.45]{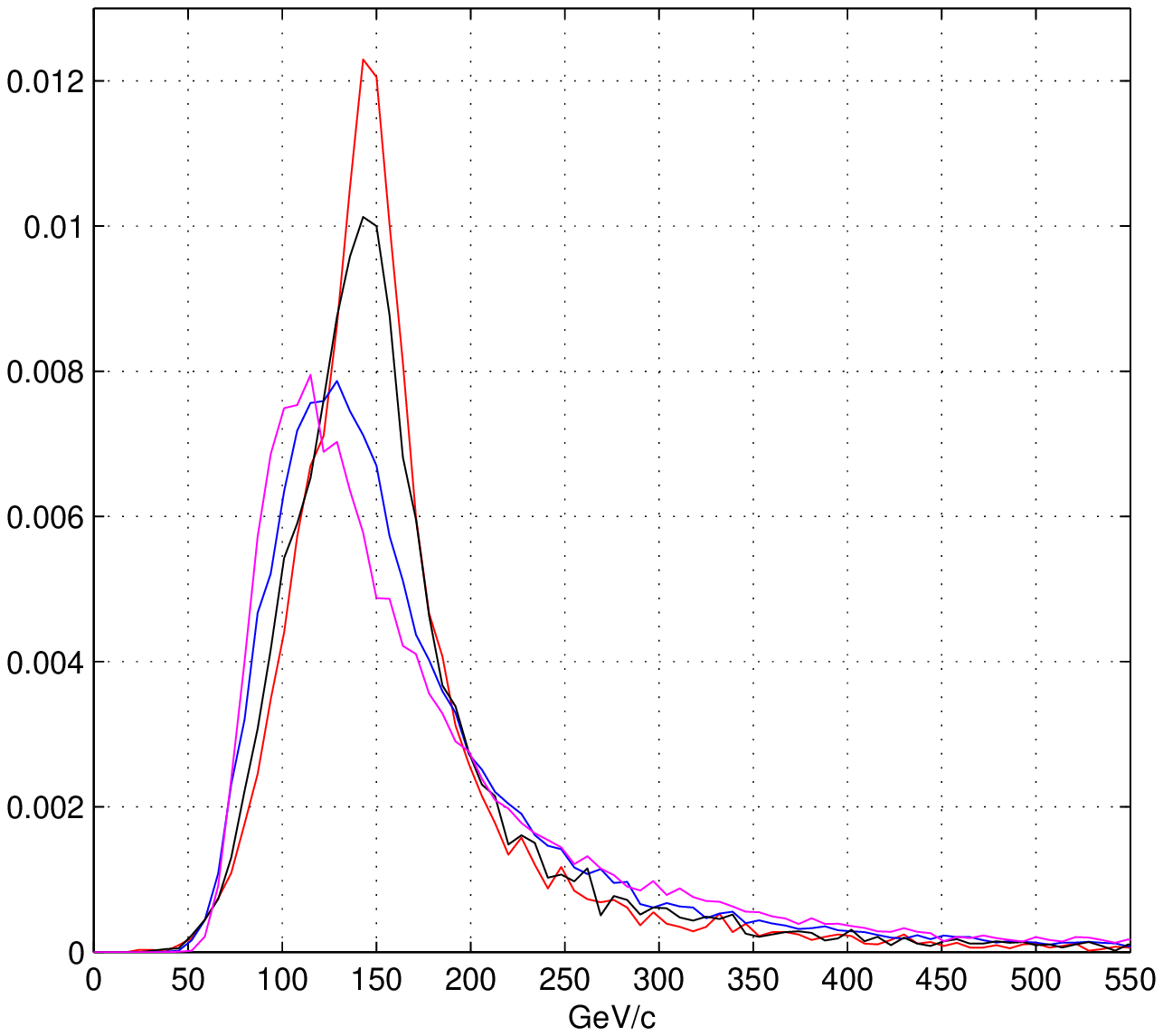}
\caption{\em Left plot. Distributions of the curvature in
$(GeV/c)^{-1}$ (the $\alpha$ parameters) MLE (red), schematic model $\sigma_{eff}(i)$ (black),
hit position $\eta_2$ (blue), hit position COG$_2$ (magenta).
Right plot:reconstructed momentum distributions.
}\label{fig:figure04}
\end{center}
\end{figure}

Figure~\ref{fig:figure03} illustrates the distributions of figure~\ref{fig:figure01}
for this type of detector, that, as discussed above, are much
lower than those of figure~\ref{fig:figure01}.
The true residuals of the four methods of track reconstruction in the left side,
and the true residuals of the least squares compared
with the position algorithms $\eta_2$ and COG$_2$.
The highest distribution of the true residuals is the MLE, followed by
the schematic model with effective $\sigma_{eff}(i)$ for each hit, the
least squares with the $\eta_2$ position algorithm, that even here is better
than the true residuals of the least squares with COG$_2$ as position algorithm.
As illustrated in the right side of figure~\ref{fig:figure03}, the data
redundancy for the least squares with the $\eta_2$-positions
is unable to improve its own
hit error PDF. The $\eta_2$ hit error
PDF is drastically better than that of
the true residuals for the fit and is the highest
distribution of plot.
As in figure~\ref{fig:figure02}, the COG$_2$ position
error PDF has two maxima with a separation four times
greater than the case of the floating strip side.
The COG systematic error is positive for the first
half of the strip and negative in the second half,
this difference of sign combines with the noise
random error to round the two maxima. The fit,
based on the COG$_2$ positions, has a good probability
to partially average these sign differences
and produce a single wide maximum. The momentum
distributions given by the four fits are reported
in figure~\ref{fig:figure04}.
The results of our MLE and the weighted least squares
are drastically better than the standard least squares.
Now the shift of the maxima for the momentum distributions
of two lowest distributions are more evident than in
figure~\ref{fig:figure02}.

\section{Discussion}

An easy comparison among the different fits is somewhat
difficult. Often an effective variance (or a standard deviation) of the PDF
is extracted interpolating a gaussian function on non
gaussian distribution to avoid the effects of the tails.
But even the variance itself is not free from arbitrariness
as often stated by Gauss in ref.~\cite{gauss}.
Similarly the full width at half maximum,
as suggested in ref.~\cite{particle_group}, does not characterize
very well non-gaussian distributions.
In any case the differences of the chosen parameters
are not of easy interpretation
from the physical point of view, and it is the physical point
of view our essential interest.
Here we will use two very "precious" physical resources as
measuring tools to establish a comparison: the magnetic field and the
signal-to-noise ratio. The simulated magnetic field intensity
and the signal-to-noise ratio are increased in the $\eta_2$
and COG$_2$ least squares reconstructions to overlap our
best distributions (the red lines). This increase is the
relative gain in resolution.

\subsection{Increasing the magnetic field}

\begin{figure}[ht]
\begin{center}
\includegraphics[scale=0.5]{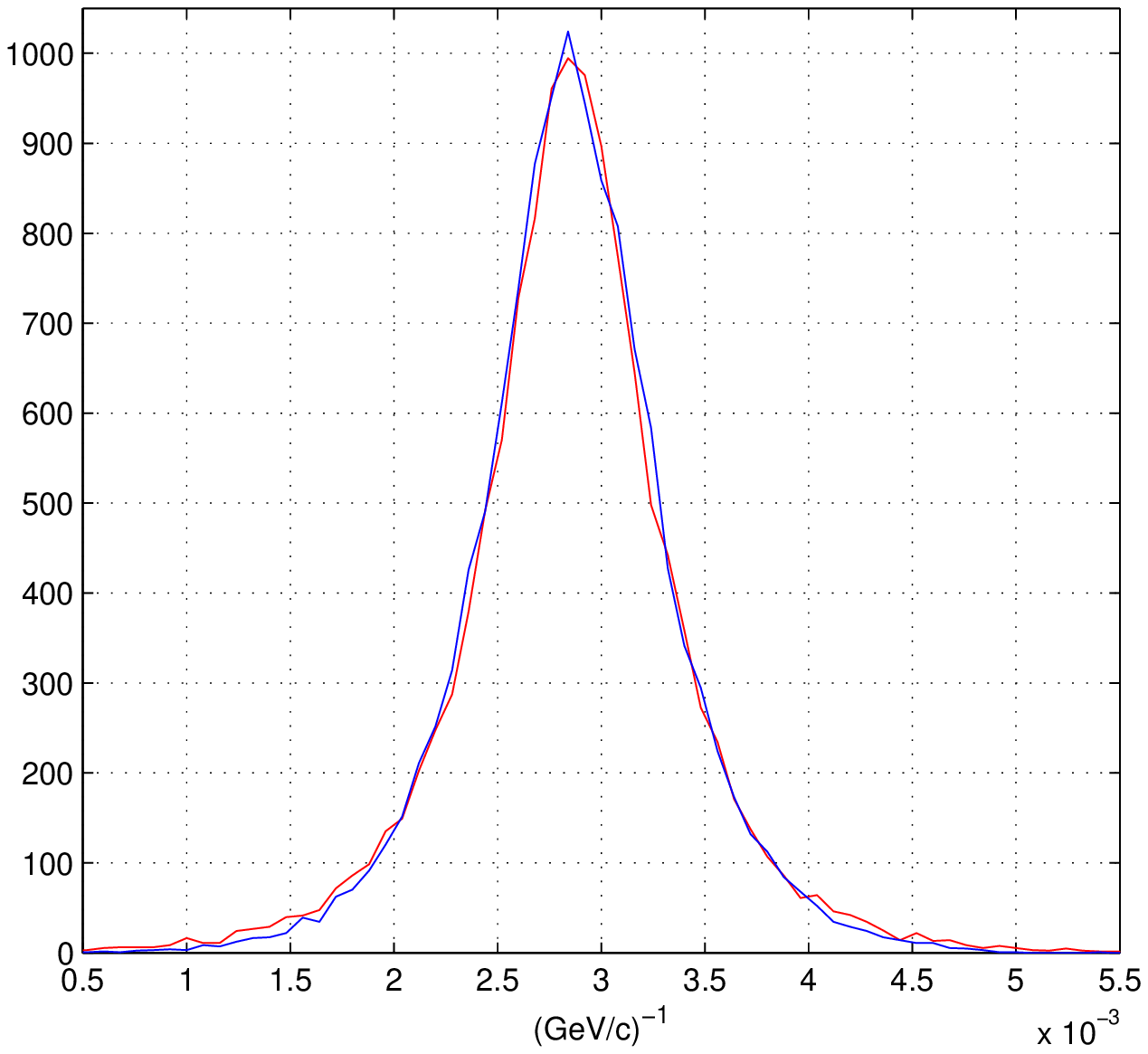}
\includegraphics[scale=0.5]{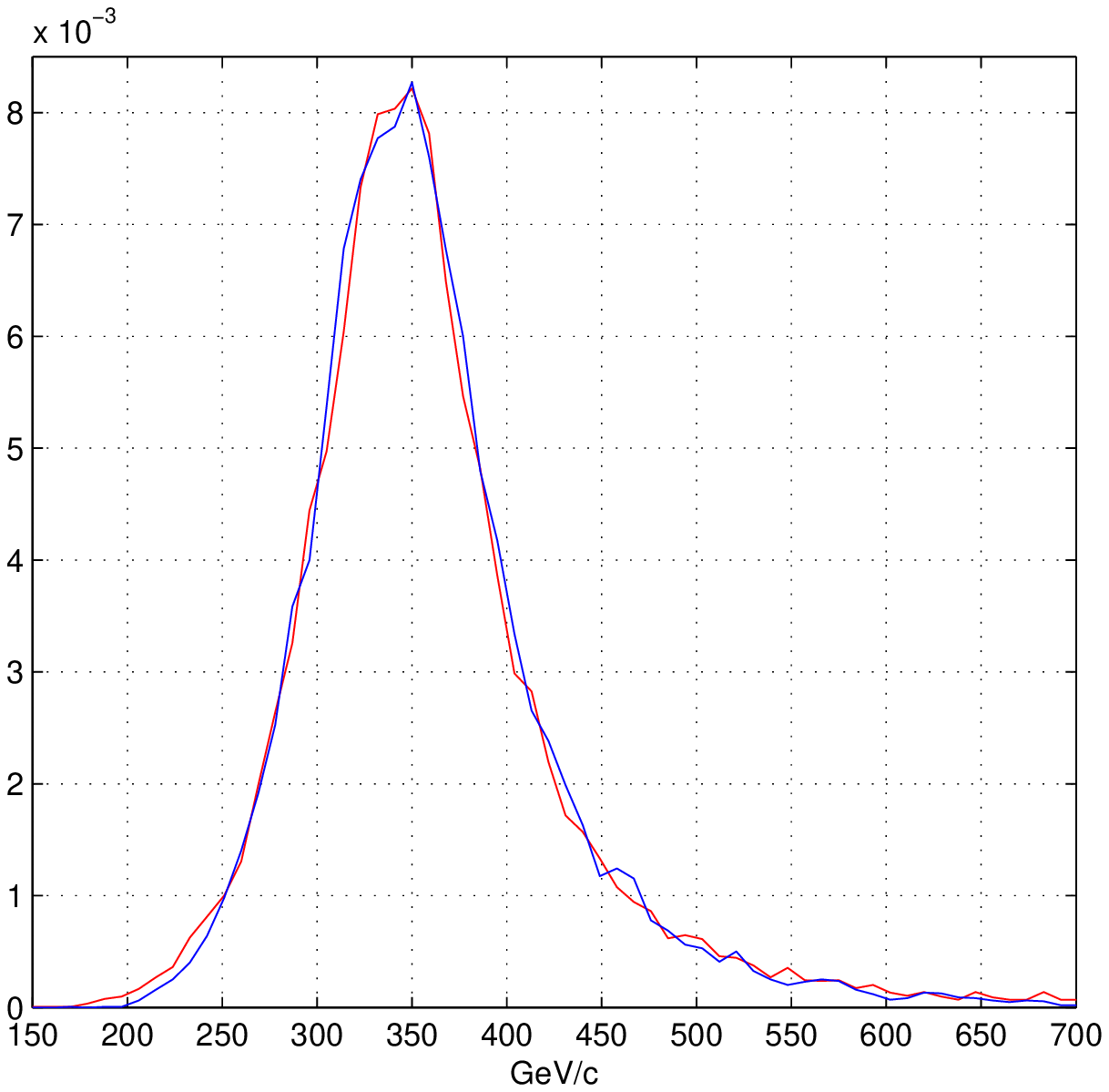}
\includegraphics[scale=0.5]{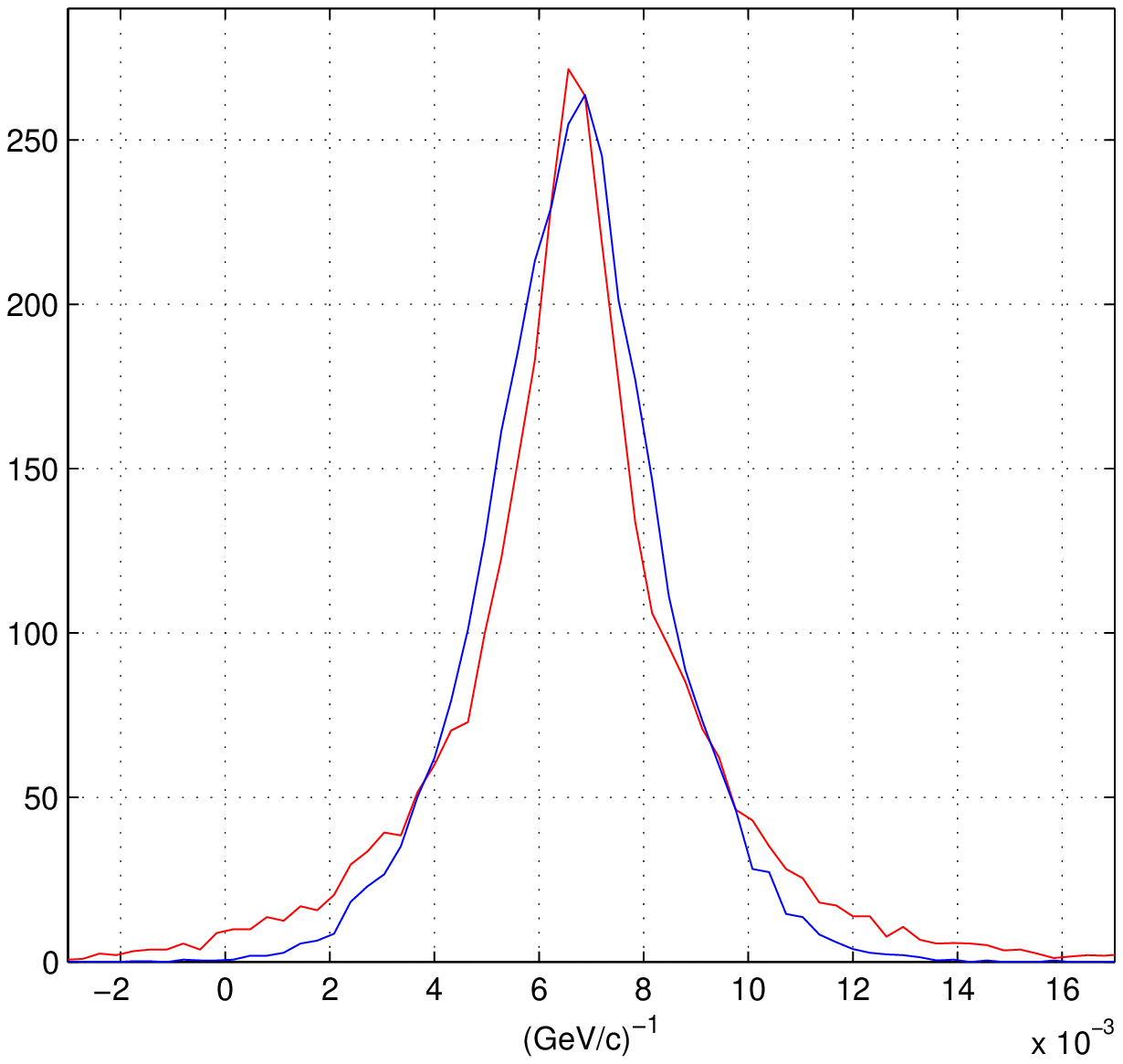}
\includegraphics[scale=0.5]{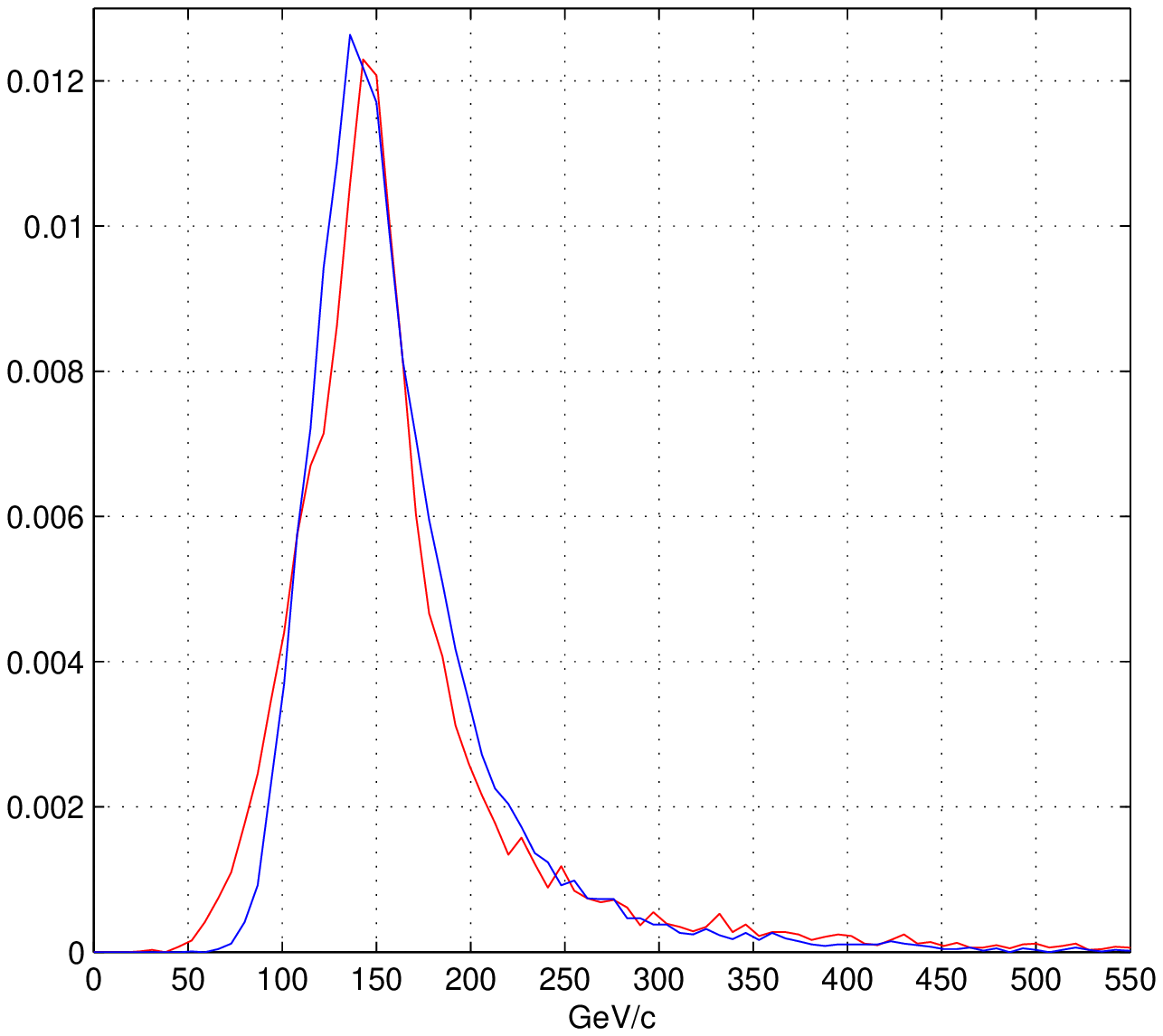}
\caption{\em Top plots. Floating strip side, Curvature in
$(GeV/c)^{-1}$ and momentum distributions for our MLE (red lines)
and $\eta_2$ least squares with a magnetic field
increased by a factor 1.5. Bottom plots. Noisy normal strip
side, here the $\eta_2$ least square has a magnetic
field 1.8 times greater than the MLE.
}\label{fig:figure05}
\end{center}
\end{figure}

With fixed momentum, the magnetic field is increased in
the fit for the $\eta_2$ least squares.
The upper sectors of figure~\ref{fig:figure05} illustrate
these results for the low noise side and the overlaps
with our MLE. The red lines are identical to those
of figure~\ref{fig:figure02}, the blue line are the
$\eta_2$ least squares for tracks with a magnetic field 1.5 times
higher. The plots for the COG$_2$ least squares are not
reported to render easily legible the figures, in this
case the magnetic field must be increased of  factor
1.8  to overlap our red lines.
The lowest sectors of figure~\ref{fig:figure05} report
the noisy normal strip side, the red lines are identical
to those of figure~\ref{fig:figure04}, the blue lines
($\eta_2$ least squares) overlap the red lines with
a magnetic field increased of 1.8 times.
Even here, the overlaps with the COG$_2$ least squares
are not reported, now, to obtain  reasonable
overlaps, the magnetic field must be doubled. Clearly,
the increment of the magnetic field produces other slight
differences in the tails of the distributions, but
the main results are well evidenced.

\subsection{Increasing the signal-to-noise ratio}

Let us discuss now the effects of the increase of the
signal-to-noise ratio. These modifications reproduce
in part the results of the magnetic increase, but
it is different in other parts. In fact, higher values  of
the magnetic field do not modify the form of the
distributions of the curvature $\alpha$ (with dimension ${legth}^{-1}$),
the curvature distributions translate toward
higher values for a reduction of the track radius.
The dimensional transformations to $(\mathrm{GeV/c})^{-1}$
and $\mathrm{GeV/c}$ have the magnetic field intensity as
scaling factor and move the distributions to
the forms of figure~\ref{fig:figure05} even for the
COG$_2$ case.
But, as for the $\alpha$  parameters
(with dimension ${length}^{-1}$), the distributions of
the true residuals of figure~\ref{fig:figure01} and
figure~\ref{fig:figure03} are not influenced by the
increase of the magnetic field, a part small effects on the
Lorentz angle.

Keeping the magnetic field to $0.44\, T$, the increase of
the signal-to-noise ratio modifies the
distributions of the $\alpha$ parameters and
those of the true residuals and can bring them
to overlap our red distributions.
In our simulations, the increase of the signal-to-noise ratio is accomplished scaling
the amplitude of the random noise,
added to the strip signals, and rerunning the
least squares fits. The plots of these results
for the momentum (in $\mathrm{GeV/c}$)
and curvature (in $(\mathrm{GeV/c})^{-1}$) are
not reported because they are practically identical to those
of figure~\ref{fig:figure05}. For the low noise side,
the  gaussian strip noise of $\sigma=4$ ADC counts
must be reduced to 2.5 ADC counts,
increasing the signal-to-noise ratio by a factor 1.6
to a huge most probable value of $56.8$.
Now, even the blue line in figure~\ref{fig:figure01} of the true
residuals reaches the red line, but the redundancy
continues to be unable to move the true residual
distribution beyond that of the new $\eta_2$ errors.
A special mention must be devoted to the COG$_2$
distributions, they remain essentially unchanged in any plot for
a (any) reduction of the strip noise. The COG$_2$
error distribution of figure~\ref{fig:figure01}
(the green line) shows a slight modification becoming
less rounded for the hard presence
of the systematic error, unaffected by the random noise.

For the higher noise side of the detector, a
reduction to less of one half of the
noise (from $8$ to $3.6\ \ \mathrm{ADC}$-counts) is
required to reproduce the low parts of figure~\ref{fig:figure05}.
With this doubling (2.2) of the signal-to-noise
ratio, the blue line (the $\eta_2$ best fit) overlaps
the red line in figure~\ref{fig:figure03} increasing the most probable
signal-to-noise ratio to 40.5.
Similarly to the other side, no noise reduction moves the
distributions given by the COG$_2$ least-squares.
As just said above, the absence of response to the
noise  reduction is due to the COG$_2$ systematic error
of ref.~\cite{landi01}. This fact renders of weak
relevance the efforts to improve the signal-to-noise
ratio in the detectors if the positioning algorithm
remains the uncorrected COG. Unpleasant effects, due
to the neglect of the systematic errors, were forecasted long
time ago in the Gauss papers~\cite{gauss}.
On the other side, this positioning method is
stable respect to a decrease of the signal-to-noise
ratio due to ageing or radiation damages,
as far as the COG systematic error continues to dominate.

A further point requires an explanation, i.e. the huge
difference between the curvature PDF
for the $\eta_2$ least squares in the low part
figure~\ref{fig:figure05} and the blue line of figure~\ref{fig:figure02}.
Now the two detector types have very similar
signal-to-noise ratio (3.6 ADC counts in the first and
4 ADC counts in the second), but, to reach the overlap
of the two PDFs, another factor of around 2.4 is
needed. This factor is too large to be due to the $20\,\%$
of differences between the sizes of their strips. The
main difference must be due to the beneficial charge
sharing of the floating strip
and the nearing of its detector architecture to the
ideal detector defined in ref.~\cite{landi01}.
Thus, detectors without this charge spreading mechanism
are evidently disadvantaged for momentum measurements.

\subsection{Different numbers of detecting layers}

\begin{figure}[ht]
\begin{center}
\includegraphics[scale=0.55]{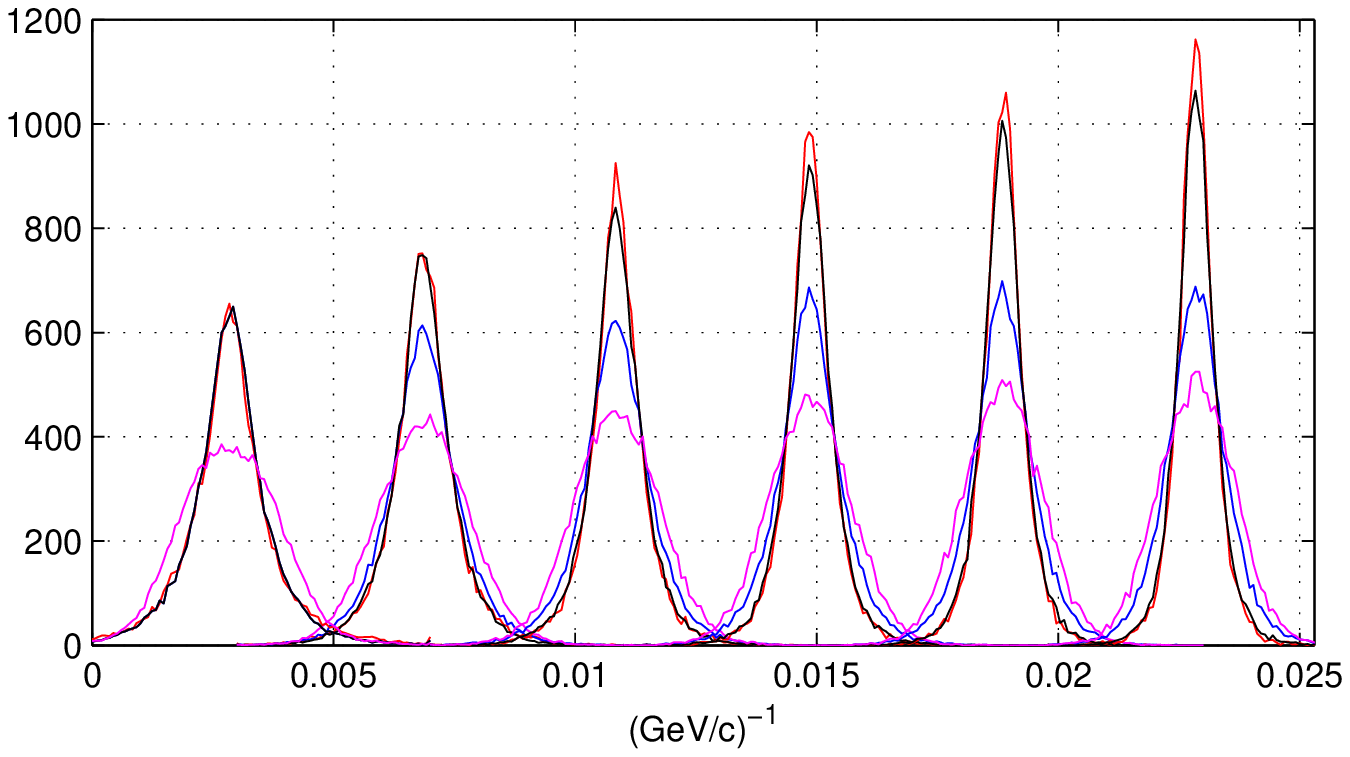}
\includegraphics[scale=0.55]{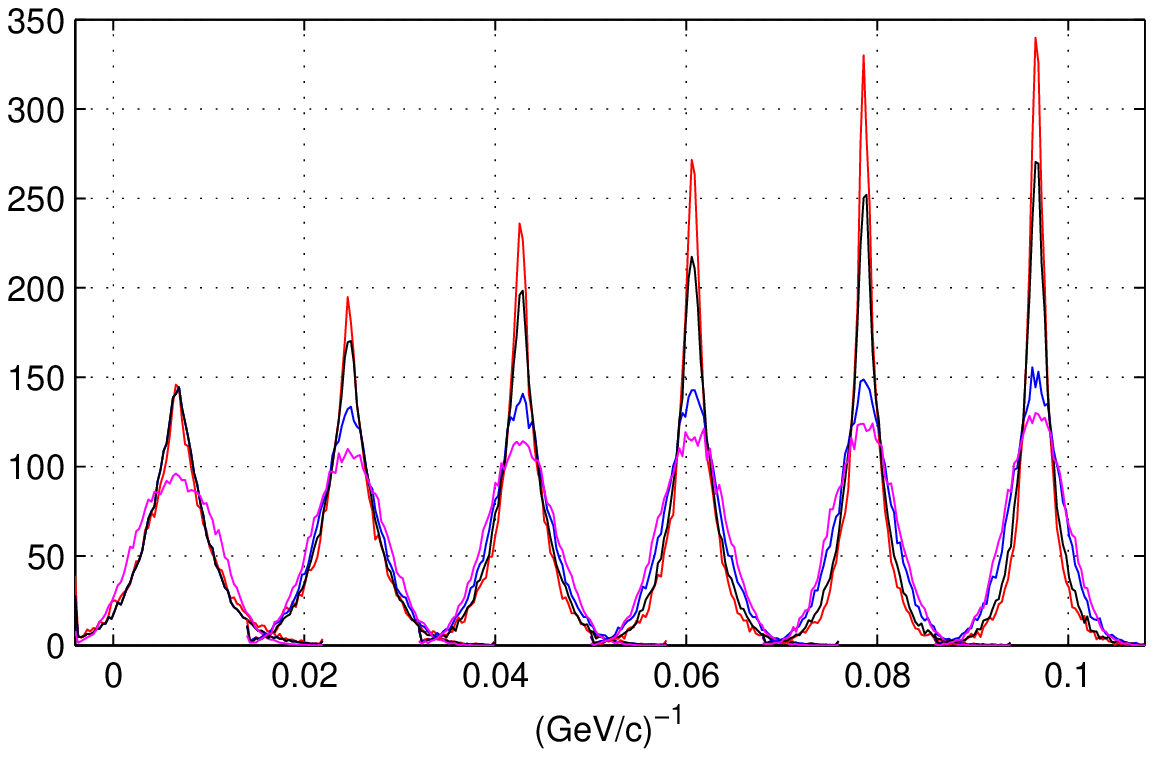}
\caption{\em Left plot. Distributions of the curvature in
$(GeV/c)^{-1}$ (the $\alpha$ parameters) for the low noise detector
and increasing the number of detecting layers. The starting one has
three layers (zero degree of freedom), the second has four layers (1 degree
of freedom) and is shifted by a fixed step to avoid the overlap with the first one.
Similarly for the other that have five, six, seven and eight detecting layers.
The right plot is similar for the higher noise side.
}\label{fig:figure06}
\end{center}
\end{figure}

Another important result of this new fitting method is the rapid grow
of the momentum resolution with the number of detecting layers.
The least squares demonstrations assume the identity of the variance of the
data to be fitted. This assumption eliminates any statistical difference among
the data and the resolution grows with the square root of the number of independent
measurements, detecting layers in our case. On the contrary our method
distinguishes the hit properties, and the quality of the track
reconstruction is controlled by the number
of excellent or good hits contained in its path. Thus we can expect that
the resolution increases with the number of detecting layers, as the
probability of excellent or good hits. To test this effect we report
in figure~\ref{fig:figure06} the grow in resolution for the curvature PDFs.

The PDFs for the six layers set up are those of figure~\ref{fig:figure02}
and of figure~\ref{fig:figure04}.
A very rapid increase, very near to a linear grow (as N),  is observed for the
maxima  of the PDFs of our two methods. (For normalized PDFs the maxima are
essentially proportional to the inverse of the full-width-at-half-maximum
that is often defined as resolution.)
The two standard least squares have the typical slow
grow of their maxima (as $\sqrt{\mathrm{N}}$).
In the case of three
layers there is a coincidence of our two methods with the least squares with the
$\eta_2$ positioning algorithm. This coincidence is due to the absence of any
redundancy (zero degree of freedom) and to the best quality of the $\eta_2$
position algorithm respect to COG$_2$. The resolution of the MLE and of the schematic
model in the four layers set up results better than the resolution of the six layers
set up for the $\eta_2$-least squares. Thus, if this resolution suffices, two detector
layers could be eliminated with important saving of weight and energy consumption that are
critical parameters for a satellite experiment. But the four layer resolution is better than
the seven and eight layers resolution of the $\eta_2$-algorithm in each of the two detector
types. Thus, important reductions of tracker complexity are possible with this
new fitting method.


\subsection{Momentum resolution for a track selection }

Up to now, we explored the momentum resolution for a generic track sample without any
special attention to the hit quality. All the generated tracks are reconstructed identically.
But our effective variance allow us to select subsets of tracks containing good or excellent hits
and to obtain much better momentum resolutions. The numbers of track collected by the running
experiments are enormous. The LHC-experiments report track numbers around $10^{12}$ per year,
thus subset of data with higher resolution could be relevant for their tasks.

The hit selections are defined in figure~\ref{fig:figure0a}, the two horizontal lines
isolate the (so called) good hits. The hits below the lowest line are defined
excellent hits. It is evident that these are somewhat arbitrary definitions
chosen to obtain easy track selections and, with a rough tuning, to have large
increases of momentum resolution.

\begin{figure}[ht]
\begin{center}
\includegraphics[scale=0.51]{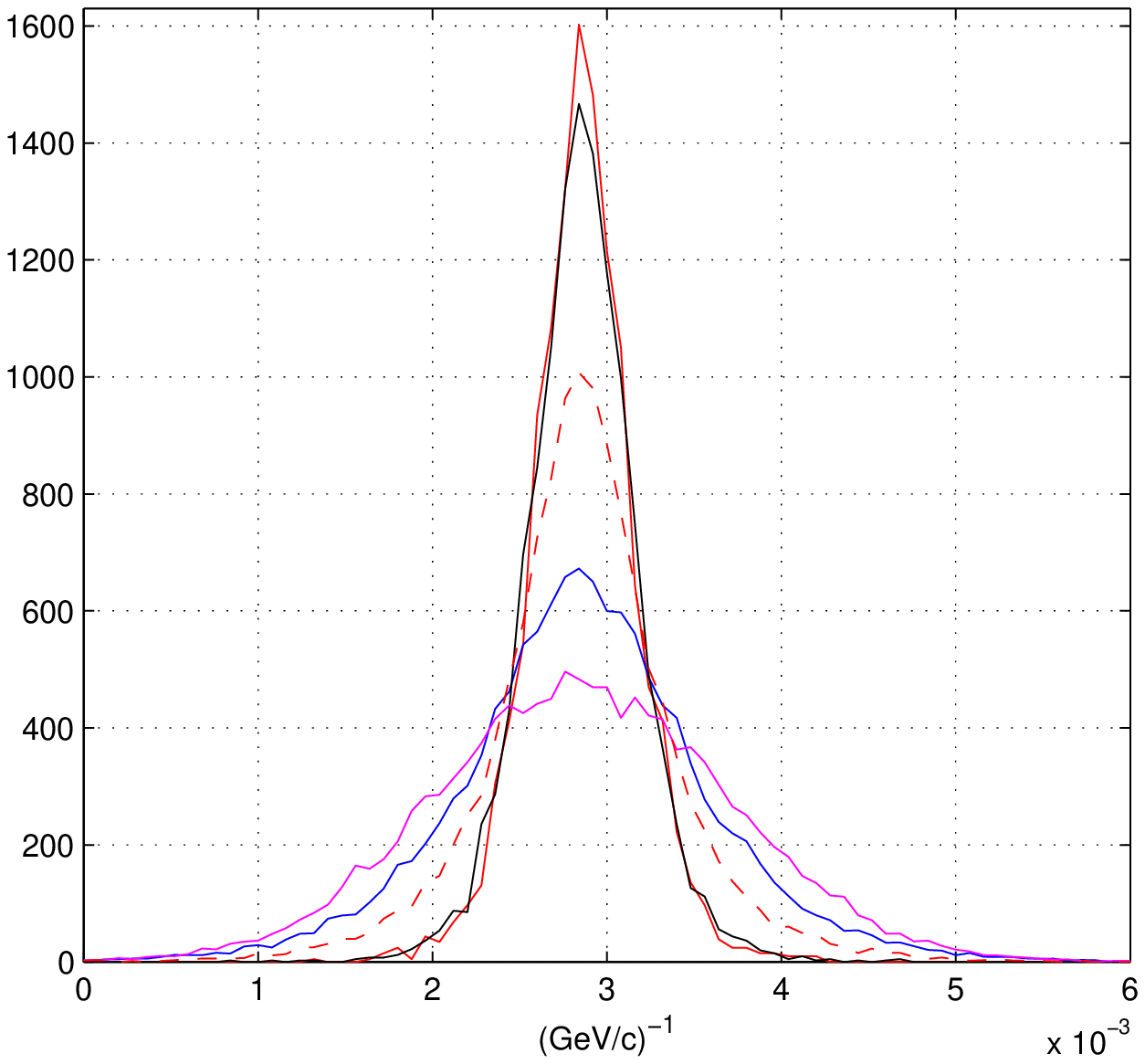}
\includegraphics[scale=0.51]{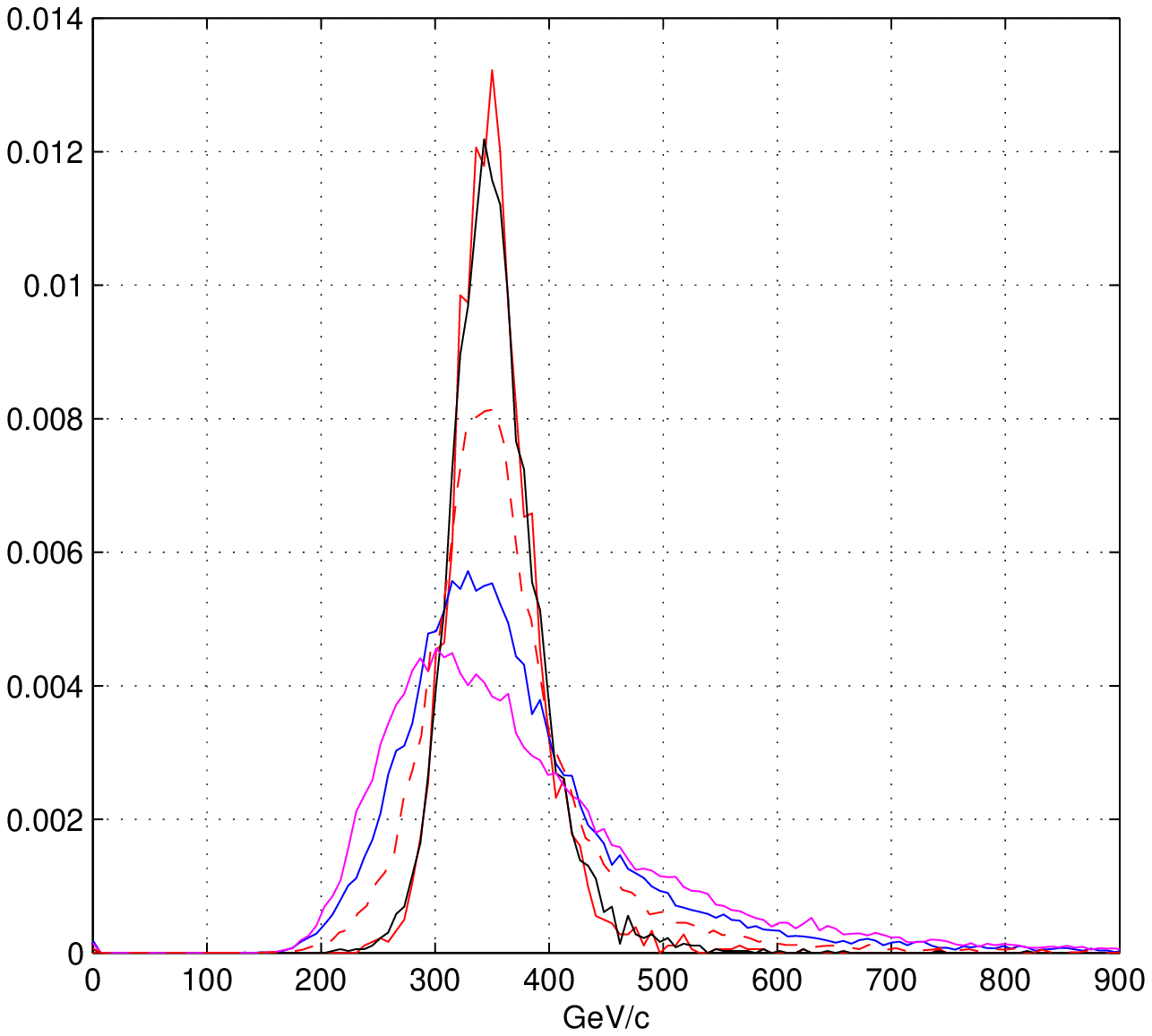}
\caption{\em Left plot. Distributions of the curvature in
$(GeV/c)^{-1}$ (the $\alpha$ parameters) for the low noise detector
for a selection of two excellent and three good hits. The two highest
distributions are the MLE and schematic model. The two lowest distributions
and the dashed one are those of figure3. To the left the momentum distributions
}\label{fig:figure08}
\end{center}
\end{figure}

The two last figures illustrate the momentum resolutions of a track selections.
For the floating strip side, we select tracks with two excellent
and three good hits. The $16.1\%$ of all the tracks has this hit combination,
and a very well defined momentum and curvature,
drastically better than that without hit selection as illustrated in figure~\ref{fig:figure08}.

\begin{figure}[ht]
\begin{center}
\includegraphics[scale=0.5]{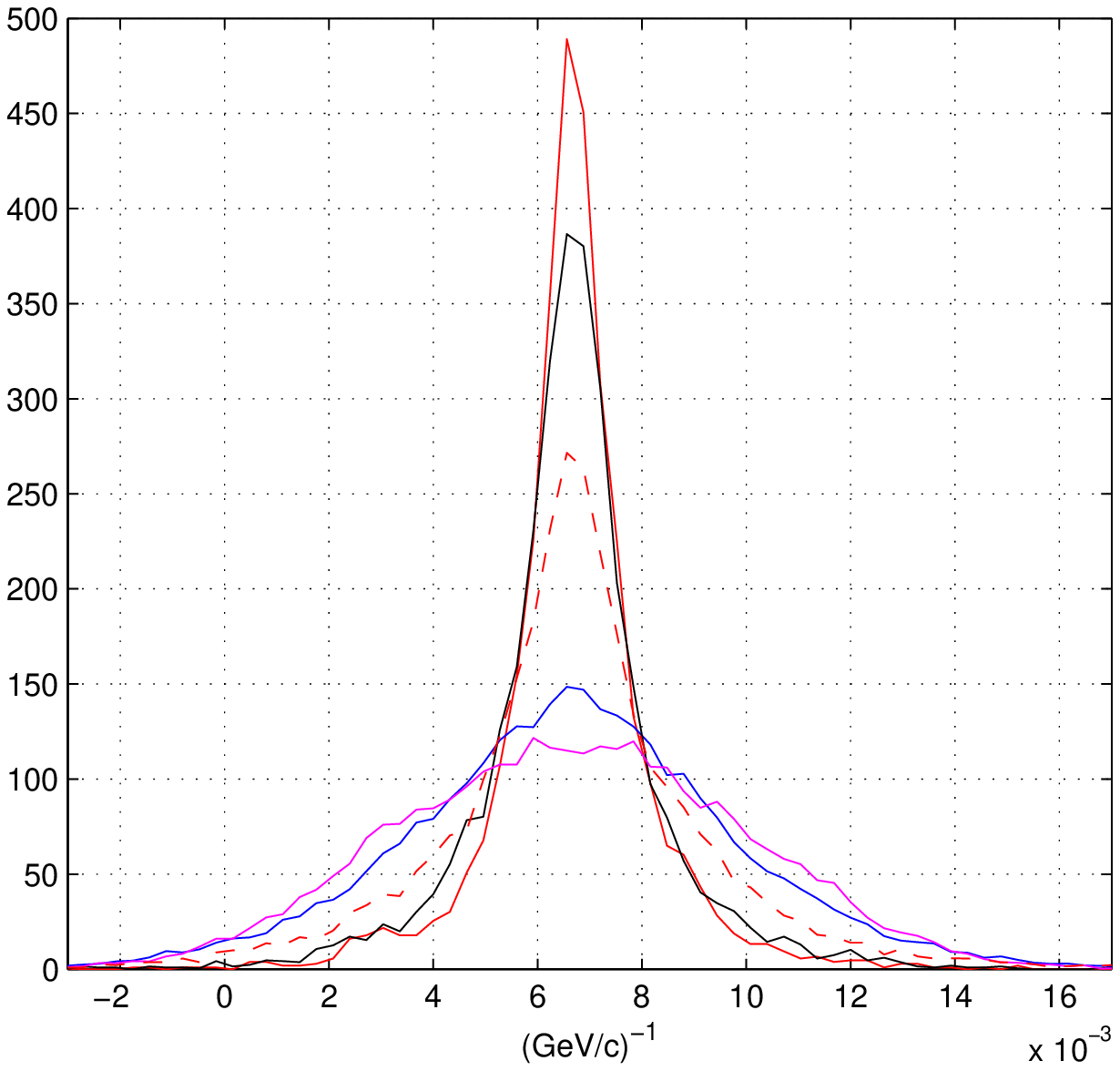}
\includegraphics[scale=0.5]{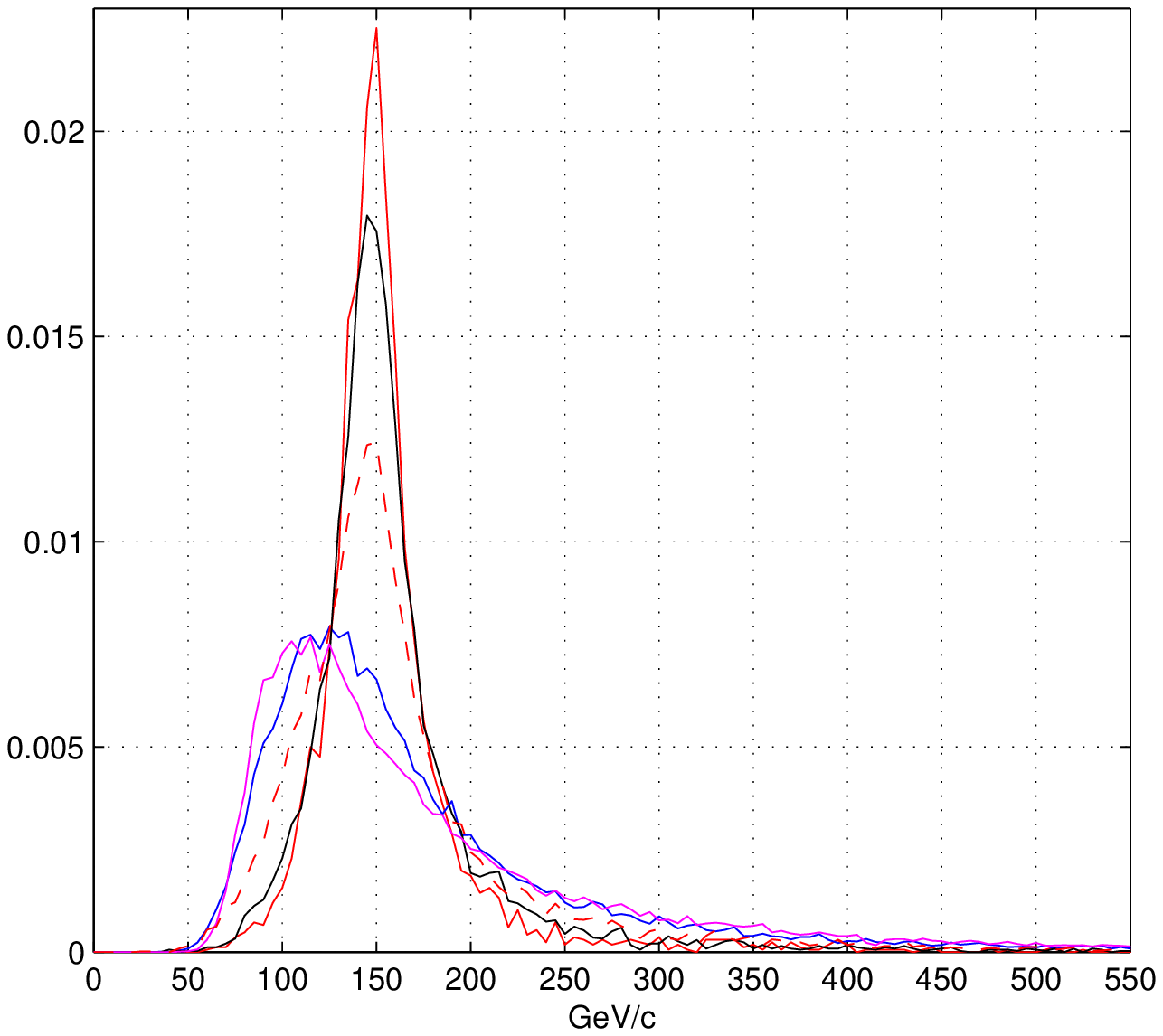}
\caption{\em Left plot. Distributions of the curvature in
$(GeV/c)^{-1}$ (the $\alpha$ parameters) for the high noise detector
for a selection of two excellent hits. The two highest
distributions are the MLE and schematic model. The two lowest distributions
and the dashed one are those of figure3. To the left the momentum distributions
}\label{fig:figure09}
\end{center}
\end{figure}

Identically we can proceed with the high noise side. Now we select tracks with
two excellent hits and four other hits of random type, $25\%$ of tracks has this hit
combination. Even in this case the momentum resolution has a great improvement
respect to the case without the hit selection. Figure~\ref{fig:figure09} shows this
comparison.

\subsection{A gift of heteroscedasticity}

The large variations of the effective variances observed in figure~\ref{fig:figure0a}
have effects even in the standard least squares. The heteroscedasticity
couples the track variance to the parameter distributions. Tracks, richer
of good or excellent hits (as we defined above), have an higher probability to
have lower residual variance. Obviously if the hit reconstruction is free from systematic errors.
The $\eta_2$-algorithm has this property. Hence, a track selection
with the lowest values of the track residual variance (whose PDF is very different from the
$\chi^2$-PDF) has the distribution of the reconstructed momenta overlapping our best PDFs
of figure~\ref{fig:figure02} and of figure~\ref{fig:figure03}. The price is an efficiency
below $25\%$ but these improvements can be obtained without any modifications of the fitting
methods. For the COG$_2$ hit positions, the track residual variance has PDF
near to a $\chi^2$, but no gift from the heteroscedasticity. The COG$_2$ positioning is dominated by its
systematic error that destroys any correlation with good hits.

\section{Conclusions}

The hit heteroscedasticity of a tracker system is
carefully explored for the momentum reconstruction
of minimum ionizing particles
in a simulated uniform magnetic field of $0.44\, T$.
Other important effects, as $\delta$-rays, multiple scattering
(negligible at high momenta), energy loss,
etc., are explicitly excluded in the simulations,
focusing on the differences among various fitting methods.
A simpler and more direct method for the calculation
of the probability density functions
is illustrated, an useful simplified form, suitable
for these applications, is reported.
Even if very synthetic this development completes the
other part illustrated in a previous publication,
and enables the replication of all the published results.
The data used for the simulations come from a test
beam with double sided detector,
this allows the extraction of realistic parameters
for two type of detectors and use each of them to
simulate the signals of charged particles in a low magnetic field.
The higher noise side has  signal properties
very similar to single sided
detectors frequently used in running trackers.
The realistic probability distributions  are applied in two form: complete
in the Maximum Likelihood Evaluation or schematic as a weight
parameters $1/\sigma_{eff}(i)^2$.
Each one of these two form is very effective for the momentum reconstructions
respect to the two other type of track
fitting i.e. the least squares with $\eta_2$ as position
algorithm and with the two strip Center of Gravity as position algorithm.
To establish a comparison among these fitting methods,
the usual method of comparing standard deviations or the
full width at half maximum is abandoned. The heavy tails of the
distributions give unrealistic contributions
to the standard deviation.  Instead, two different
tracker properties are modified to reach
an overlap among the fit outputs: the magnetic
field and the signal-to-noise ratio.
To reach the overlap of the two standard fits with the
best momentum distributions, the magnetic field must
grow by a factor 1.5 for the $\eta_2$
positioning, and 1.8 for  the two strip Center of Gravity in the low
noise side and 1.8 and 2 for the higher noise side.
The increase of signal-to-noise ratio is effective
only for the $\eta_2$ position algorithm, the overlaps are obtained
with  factors 1.6 and 2.2  for the two detector sides.
Any increase of the signal-to-noise ratio has no effect
in the least squares based on two strip Center of Gravity. The
intrinsic systematic error of the Center of Gravity
as position algorithm survives untouched to any reduction of
the detector random noise biasing any position measurement.
An evident difference
in the resolution of the momentum reconstruction is
observed in the noisy detector side respect to the
floating strip side. Even if the noise is reduced to
that of the other side, the momentum resolution remains
drastically lower. The effect
of the floating strips are very beneficial for momentum
resolution. The nearing to the ideal detector, as defined
in a our previous work, attenuates the effects of the
heteroscedasticity.
The simulations are extended to the study of the
effect of the number of detection layers on momentum
resolution. It is easy to observe an almost linear
increase of resolution of the Maximum Likelihood Evaluation and of the
schematic model, as expected, being bound the linear grow
of the probability of good or excellent hits.
Standard fits grow as the square root of the layer
number as for any linear regression model with homoscedasticity.
To test the dependence from the hit quality, a rough
definition of good and excellent hit is introduced and
tracks with a given content of good and excellent hits are selected.
These selected tracks show an additional increase of
the momentum resolution at the price of an efficiency reduction.
It must be reminded that
these are simulations and are accompanied
with the usual uncertainties of any simulation.
Assuming an optimistic view, this increase in
resolution can be spent in different ways
either for better results on running experiments
or in reducing the complexity of future experiments if the
baseline fits (almost always based on the COG
positioning) are estimated sufficient.
Further details about this method will be published
elsewhere.

\end{document}